\documentclass[]{aastex631}
\usepackage{subfigure}
\usepackage{multirow}
\usepackage{longtable}
\usepackage{amsthm,amsmath,amssymb}
\usepackage{mathrsfs}
\usepackage{threeparttable}
\usepackage{graphicx}

\shorttitle{Classification and physical characteristics analysis of Fermi-GBM GRB based on DL}
\shortauthors{Chen et al.}
\graphicspath{{./}{figures/}}

\begin{document}

\title{Classification and physical characteristics analysis of Fermi-GBM Gamma-ray bursts based on Deep-learning}

\correspondingauthor{Li Zhang}\email{lizhang@ynu.edu.cn}

\author[0000-0001-5681-6939]{Jia-Ming Chen}
\affiliation{Department of Astronomy, School of Physics and Astronomy, Key Laboratory of Astroparticle Physics of Yunnan Province, Yunnan University, Kunming 650091,China}

\author[0000-0002-3132-1507]{Ke-Rui Zhu}
\affiliation{Department of Astronomy, School of Physics and Astronomy, Key Laboratory of Astroparticle Physics of Yunnan Province, Yunnan University, Kunming 650091,China}

\author{Zhao-Yang Peng}
\affiliation{College of Physics and Electronics, Yunnan Normal University, Kunming 650500, China }

\author{Li Zhang}
\affiliation{Department of Astronomy, School of Physics and Astronomy, Key Laboratory of Astroparticle Physics of Yunnan Province, Yunnan University, Kunming 650091,China}

\begin{abstract}
The classification of Gamma-Ray Bursts  has long been an unresolved problem. Early long and short burst classification based on duration is not convincing due to the significant overlap in duration plot, which leads to different views on the classification results. We propose a new classification method based on Convolutional Neural Networks and adopt a sample including 3774 GRBs observed by Fermi-GBM to address the $T_\text{90}$ overlap problem. By using count maps that incorporate both temporal and spectral features as inputs, we successfully classify 593 overlapping events into two distinct categories, thereby refuting the existence of an intermediate GRB class. Additionally, we apply the optimal model to extract features from the count maps and visualized the extracted GRB features using the t-SNE algorithm, discovering two distinct clusters corresponding to S-type and L-type GRBs. To further investigate the physical properties of these two types of bursts, we conduct a time-integrated spectral analysis and discovered significant differences in their spectral characteristics. The analysis also show that most GRBs associated with kilonovae belong to the S-type, while those associated with supernovae are predominantly L-type, with few exceptions. Additionally, the duration characteristics of short bursts with extended emission suggest that they may manifest as either L-type or S-type GRBs. Compared to traditional classification methods (Amati and EHD methods), the new approach demonstrates significant advantages in classification accuracy and robustness without relying on redshift observations. The deep learning classification strategy proposed in this paper provides a more reliable tool for future GRB research.
\end{abstract}

\keywords{Gamma-ray bursts(629)}

\section{Introduction} \label{intro}

Gamma-Ray Bursts (GRBs) are one of the most intense explosive phenomena in the universe. The duration of GRBs shows a bimodal distribution, which was established early on \citep{1981Ap&SS..80....3M,1984Natur.308..434N}. \cite{1993ApJ...413L.101K} analyzed  GRBs observed by CGRO/BATSE and introduced the concept of $T_\text{90}$, the time needed to detect 90\% of the total GRB fluence. They found that  $T_\text{90}$ has a bimodal distribution with a dividing point at 2 seconds (s). This discovery became the basis for classifying GRBs into long GRBs (LGRBs, $T_\text{90}>2$ s) and short GRBs (SGRBs, $T_\text{90}<2$ s). Additionally, \cite{1993ApJ...413L.101K} found that the energy spectra of LGRBs are softer, while those of SGRBs are harder.

LGRBs are thought to be formed by the collapse of massive star cores \citep{1993ApJ...405..273W,1998ApJ...494L..45P,2006RPPh...69.2259M,2006ARA&A..44..507W}, a theory supported by direct observational evidence that LGRBs are associated with broad-lined Type Ic supernovae \citep{1998Natur.395..670G,1998Natur.395..663K,2003ApJ...591L..17S,2017A&A...605A.107C}. In contrast, SGRBs are thought to originate from the merger of compact objects  \citep{1986ApJ...308L..43P,1989Natur.340..126E,1992ApJ...397..570M}, such as two neutron stars (NS-NS) or a neutron star and a black hole (NS-BH). This view is confirmed by the multi-messenger observations of the neutron star merger event GW170817/GRB 170817A \cite{2017ApJ...848L..12A,2017ApJ...848L..14G}.

Although the dichotomous classification of GRBs initially achieved some success, advancements in observational technology have revealed significant limitations in this method. First, the duration distributions of LGRBs and SGRBs significantly overlap \citep{2016ApJS..227....7L}, and the duration itself depends on the energy band being measured \citep{2006Natur.444.1047F,2009ApJ...703.1696Z,2010ApJ...725.1955V,2013MNRAS.430..163Q,2013ApJ...764..179B,2016SSRv..202....3Z}. Additionally, the existence of ``hybrid" GRBs challenges the practice of classifying GRBs based solely on duration \citep{2009Natur.461.1254T,2009A&A...507L..45A,2011A&A...531L...6N,2022JApA...43...39D,2023MNRAS.522.5204B}, indicating the need for more complex classification criteria. For example, GRB 200826A is a short-duration GRB associated with a stellar core collapse origin \citep{2021NatAs...5..917A,2021NatAs...5..911Z,2022ApJ...932....1R}, whereas GRB 211211A \citep{2022Natur.612..232Y,2022Natur.612..228T,2022Natur.612..223R,2023NatAs...7...67G} and GRB 230307A \citep{2024ApJ...962L..27D,2024Natur.626..737L,2024Natur.626..742Y,2024ApJ...964L...9W,2024ApJ...969...26P} are long-duration GRBs that are believed to potentially originate from the merger of compact stars and are associated with kilonovae. Furthermore, some long-duration GRBs, such as GRB 060505 \citep{2007ApJ...662.1129O,2007ApJ...667L.121L,2008ApJ...677L..85M} and GRB 060614 \citep{2006Natur.444.1053G,2006Natur.444.1044G,2007ApJ...655L..25Z}, were not accompanied by observed supernovae \citep{2006Natur.444.1047F}. Therefore, $T_\text{90}$ alone cannot determine the progenitor type associated with a GRB, and reliably identifying the progenitor requires the observation of supernova or kilonova emissions. The traditional view suggests that LGRBs originate from the collapse of massive stars, implying that their formation rate should align with the cosmic star formation rate, whereas the formation rate of short SGRBs should lag behind.  However, \cite{2024ApJ...963L..12P} have found that the formation rate of low-redshift LGRBs is significantly higher than the cosmic star formation rate and is very similar to the formation rate of SGRBs. Additionally, some low-redshift LGRBs are associated with kilonovae. These findings challenge the traditional view that LGRBs primarily originate from the collapse of massive stars.

$T_\text{90}$ is an important characteristic of prompt radiation and has been extensively studied in the literature. \cite{1998ApJ...508..757H} first proposed the existence of a third category of intermediate GRBs by fitting a Gaussian distribution to the log($T_\text{90}$) distribution of 797 GRBs in the BATSE 3B catalog. This conclusion was further validated in \cite{2002A&A...392..791H} study. With the advent of the Swift era, \cite{2008A&A...489L...1H} analyzed 222 GRB samples from the Swift-BAT catalog and found similar results supporting a trimodal distribution. Subsequently, \cite{2016Ap&SS.361..155H} conducted an analysis with a larger sample and reached the same conclusion. However, in the Fermi era, \cite{2015A&A...581A..29T} studied the log($T_\text{90}$) distribution of 1566 GRBs in the Fermi-GBM catalog using \cite{1998ApJ...508..757H} criteria, but found the distribution to be essentially bimodal, with no evidence of a third category. \cite{2017Ap&SS.362...70K} also conducted similar research on the log($T_\text{90}$) distribution by comparing two (or three) log-normal distribution components. In addition to GRBs observed by BATSE 4B, Swift-BAT, and Fermi-GBM, they also included GRBs observed by BeppoSAX but were unable to reach a conclusive result regarding the existence of a third GRB category. Therefore, the nature of intermediate GRBs remains unclear and controversial.

To achieve more accurate classifications of GRBs, researchers have proposed various classification schemes. Most of these efforts are summarized in Table 1 of \cite{2022Galax..10...77S}. These attempts aim to identify GRB categories using parameters such as duration, hardness ratio, flux, and more \citep{2003ApJ...582..320H,2016MNRAS.462.3243Z,2017Ap&SS.362...70K,2022Ap&SS.367...39B,2022MNRAS.517.5770Z}. Depending on the samples, parameters, and methods used, two to five categories of GRBs can be identified. These classification methods strive to improve accuracy by incorporating multiple observational parameters, and all rely on the parameter $T_\text{90}$. The traditional classification of LGRBs and SGRBs is still mainly based on community consensus, lacking an objective classification model with minimal human intervention.

In this context, the application of machine learning (ML) techniques has gradually become a focal point in GRB classification research. ML can automatically generate results after training, without the need for manual input, helping researchers better understand the differences between LGRBs and SGRBs, and classify newly discovered GRBs. Over the past few decades, ML has been widely used in GRB research. For example, \cite{2019Ap&SS.364..105H} used principal component analysis and statistical clustering techniques to classify 801 GRB samples from the Fermi GBM catalog into one short burst category and two long burst categories distinguished by peak flux, highlighting the complexity of GRB classification. \cite{2021A&C....3400441M} used fuzzy clustering analysis to confirm three known groups and discovered five new groups, further complicating GRB classification. \cite{2022A&A...657A..13T} found through k-nearest neighbor graph classification that GRBs could be divided into two or three groups, but the results varied with increasing sample size, failing to clearly support the existence of more than three categories. These studies indicate that GRB classification is more complex and diverse than previously thought. Additionally, recent studies have employed unsupervised dimensionality reduction algorithms such as t-distributed stochastic neighbor embedding (t-SNE) and uniform manifold approximation and projection (UMAP) to analyze GRB light curves and spectra, successfully classifying GRBs \citep{2020ApJ...896L..20J,2023ApJ...945...67S,2023ApJ...951....4G,2024MNRAS.527.4272C,2024ApJ...969...88M,2024arXiv240605005D}. \cite{2023MNRAS.525.5204B} used unsupervised machine learning methods to perform cluster analysis on optical and X-ray samples of GRBs with a plateau phase, exploring various combinations of plateau phase and prompt emission parameters to identify GRB characteristics that yield distinct clustering effects. The study highlights the importance of the ``plateau phase" feature in revealing the physical mechanisms of GRBs. However, limitations in observational conditions have led to widespread missing data for the plateau phase. To address this, researchers have proposed various methods, including machine learning, to reconstruct GRB light curves. For example, \cite{2023ApJS..267...42D} used a  reconstruction method based on stochastic processes, while \cite{2024sais.conf..543S} applied a bidirectional long short-term memory network (BiLSTM) to reconstruct GRB light curves. These studies further demonstrate the potential of machine learning techniques in revealing the complex classification structure of GRBs.

However, unsupervised learning methods rely only on input feature data and do not include category labels, which limits their classification accuracy. In contrast, supervised learning methods have significant advantages in classification tasks. Supervised learning methods have been used in GRB classification research. Recently, \cite{2023ApJ...959...44L} used  eXtreme Gradient Boosting (XGBoost) classifier, a supervised machine learning method, to train a model that classified GRBs into Type I and Type II. This excluded the existence of a third intermediate GRB category based on $T_\text{90}$ duration distribution. Deep learning (DL) is a subset of unsupervised machine learning that has achieved great success in computer vision and natural language processing. It also has many applications in astronomy and astrophysics \citep{2022JApA...43...76K,2022ExA....53....1S}. Convolutional neural network (CNN) are a commonly used DL algorithm that can automatically extract features from high-dimensional data. CNN are useful in image classification because they learn intrinsic features of samples layer by layer and maintain spatial relationships between pixels. CNN have been widely used in astronomical research. For example, \cite{2022RAA....22j5007L} used  CNN algorithm to search for weak fast radio bursts in four years of observational data from the Parkes radio telescope. \cite{2024ApJS..272....4Z} used CNNs to identify GRBs and showed that DL methods can effectively and reliably distinguish GRBs from similar background images. \cite{2023ApJ...945..106P} used  CNN autoencoder to reconstruct light curves of background-only data and searched for GRB by detecting reconstruction errors that exceeded a threshold.

Supervised learning methods, with their ability to combine features and labels for classification, have shown great potential in GRB research. By applying supervised learning techniques, we can more accurately identify the true physical origins of hybrid GRBs. In this paper, we develop a CNN model to perform a classification of  GRBs, mainly focusing on $T_\text{90}$ overlapping GRBs. The structure of this paper is organized as follows: Section \ref{data} outlines the construction of the training, validation, and test datasets; Section \ref{Methodology} provides a detailed description of the model architecture and training process, as well as the t-SNE method used for dimensionality reduction and visualization; Section \ref{Result} presents the model's test results and its application in GRB classification and feature extraction; Section \ref{Discussion} discusses the classification results; and finally, Section \ref{Summary} concludes the study with a summary and conclusions. 

\section{Data and Samples \label{data}} 
In this study, we would like to classify GRBs based on the overlapping $T_\text{90}$ data. We choose a sample observed by the Fermi-GBM detector because the instrument covers a larger energy range, which has detected 3774 GRBs as of the end of May 2024. With the dataset we can construct a dataset containing training sample and classification sample to train our deep learning model.

\begin{figure}[htbp]
\centering
\subfigure[]
{
\begin{minipage}[t]{0.45\textwidth}
\centering
\includegraphics[width=\textwidth]{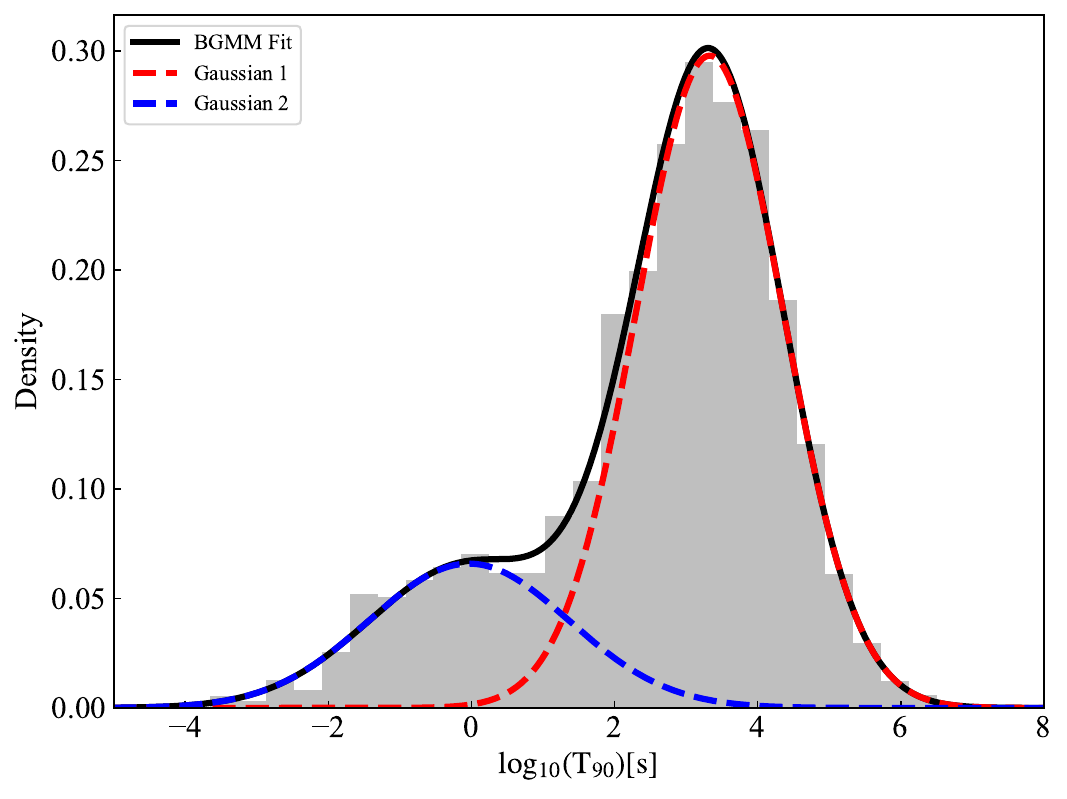}
\end{minipage}
}
\subfigure[]
{
\begin{minipage}[t]{0.45\textwidth}
\centering
\includegraphics[width=\textwidth]{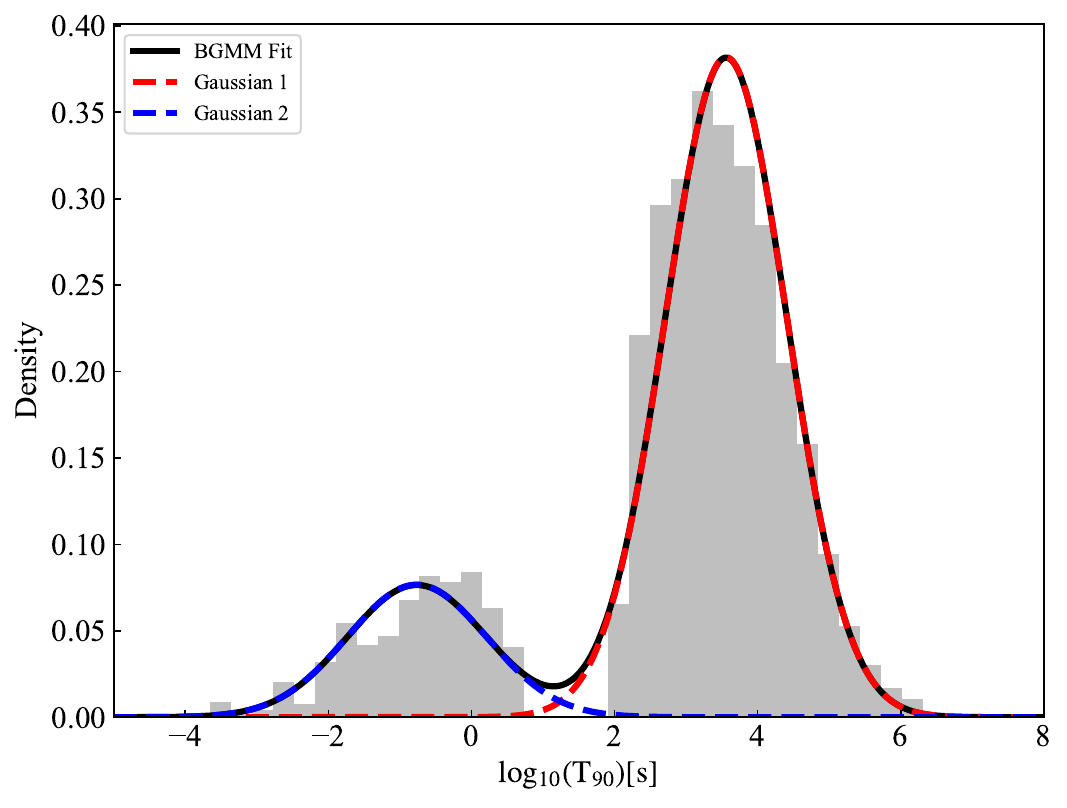}
\end{minipage}
}
\caption{BGMM fitting results for the $log_\text{10}(T_\text{90})$ data of the Fermi-GBM GRB. (a) Panel shows the BGMM fitting results for the complete dataset of 3774 GRBs. (b) Panel shows the BGMM fitting results for the remaining 3177 GRBs after removing the overlapping portion. The black curve represents the overall BGMM fit. The blue and red dashed curves represent the individual Gaussian components for short-duration and long-duration GRBs, respectively.\label{fig1}}
\end{figure}

First, to identify GRBs in the overlapping region, we use a Bayesian Gaussian Mixture Model (BGMM) to fit the $log_\text{10}(T_\text{90})$ data of the 3774 GRBs. The BGMM combines Bayesian inference and Gaussian Mixture Models (GMM), serving as a probabilistic model for clustering analysis and density estimation. Figure \ref{fig1}(a) shows the fitting results, indicating a clear bimodal Gaussian distribution in the $log_\text{10}(T_\text{90})$ data: one component corresponds to short-duration GRBs, while the other corresponds to long-duration GRBs, with an overlapping region in between. The fitting results also provide the posterior probability of each GRB belonging to each Gaussian component.

Based on these posterior probabilities, we set a threshold (0.12) to determine the GRBs belonging to the overlapping region. Using this method, we identify 594 overlapping GRBs with $T_\text{90}$ ranging from 1.792 s to 8.256 s. These overlapping GRBs constitute our classification samples. Additionally, the classification samples include three special GRBs: GRB 200826A, GRB 211211A, and GRB 230307A.

Figure \ref{fig1}(b) shows the BGMM fitting results for the remaining 3177 $log_\text{10}(T_\text{90})$ after removing the overlapping portion. It can be seen that the remaining samples can be fitted by two Gaussian components without overlap. We label the short-duration and long-duration components as SGRB and LGRB, respectively, serving as our training samples. We then split the SGRB and LGRB samples into training, validation, and test sets in a 7:2:1 ratio. The training set is used to train the model, the validation set is used to validate the model's performance after each training epoch, and the test set is used to evaluate the generalization ability of the final optimized model.

\begin{figure}[htbp]
\centering
\begin{minipage}[t]{1.0\textwidth}
\centering
\includegraphics[width=\textwidth]{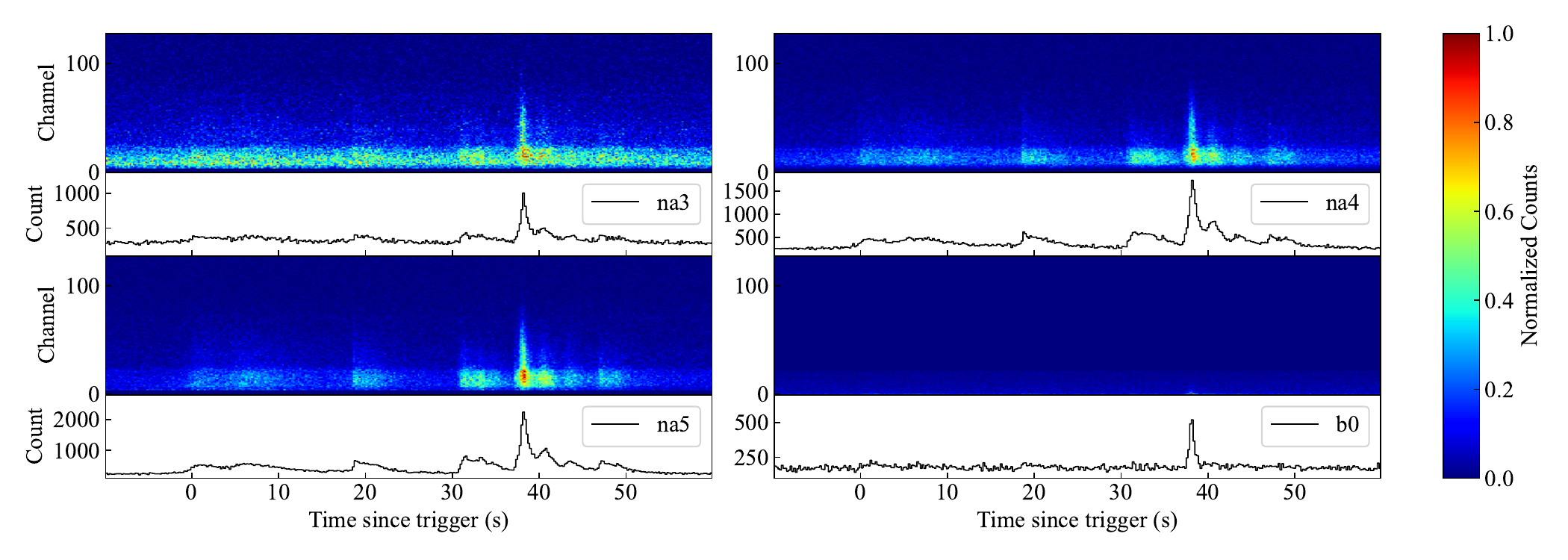}
\end{minipage}
\caption{Count maps and light curves of the GRB240205926 event recorded by four detectors (n3, n4, n5, and b0). \label{fig2}}
\end{figure}

The GRB count maps preserve both temporal and spectral information, encompassing essential details about the radiation mechanisms of GRBs. \cite{2024ApJS..272....4Z} used count maps as input to effectively distinguish GRBs from non-GRBs through CNN. To obtain the GRB count maps, we design a Python program to extract the count map for each GRB. The program uses the \texttt{GBM-Data-Tools} \citep{GbmDataTools} package to access and analyze GBM data. We use publicly available GRB Time-Tagged Event (TTE) format data\footnote{Data obtained from \url{https://heasarc.gsfc.nasa.gov/W3Browse/fermi/fermigbrst.html}}, which covers the entire $T_\text{90}$ period for each GRB event, from about 20 s before the trigger to about 300 s after the trigger. 
For each GRB, we selected the first 15 s and the last 30 s of its duration as the time range for sampling, ensuring that key emission features are captured. Additionally, to facilitate training of the machine learning model, we standardize the time length of each GRB to 512. We divide the selected time range by 512 to determine the photon count resolution (bin). Keeping the time length at 512 can effectively balance capturing the essential characteristics of GRB and managing computing resources. This setup allows us to finely resolve the temporal variations in SGRBs without imposing an excessive data processing burden. Furthermore, longer sequences could significantly increase the computational load and extend training time without yielding substantial performance gains. Standardizing the input length to 512 ensures efficient model training and prevents memory issues when handling large datasets. To eliminate amplitude differences between samples, each count map is individually normalized by rescaling to between 0 and 1. Therefore, in the count maps, the x-axis represents time with a length of 512, the y-axis represents 128 energy channels, and the color indicates the normalized photon counts (as shown in Figure \ref{fig2}).

As mentioned earlier, our sample comprises 3774 GRBs, with 3177 used for the training set. However, only 3726 GRBs successfully generate count maps. This may result in our dataset being relatively small for training a DL model. Each GRB event is detected by Fermi-GBM's 12 NaI detectors (energy range: 8 - 1000 keV) and 2 BGO detectors (energy range: 200 keV - 40 MeV), with slight variations in each event due to direction and detector response. Thus, events triggered by each individual detector are considered independent GRB events. We can treat signals from each triggering detector as independent GRB sample. Considering the weaker signals from some detectors, we select data from the three brightest NaI detectors and one BGO detector, thereby quadrupling our sample size. Figure \ref{fig2} shows the count maps of GRB 240205926 in the four detectors.

\section{Methodology \label{Methodology}} 

\subsection{Convolutional Neural Network Model \label{CNN}}
\begin{figure}[htbp]
\centering
\begin{minipage}[t]{1.0\textwidth}
\centering
\includegraphics[width=\textwidth]{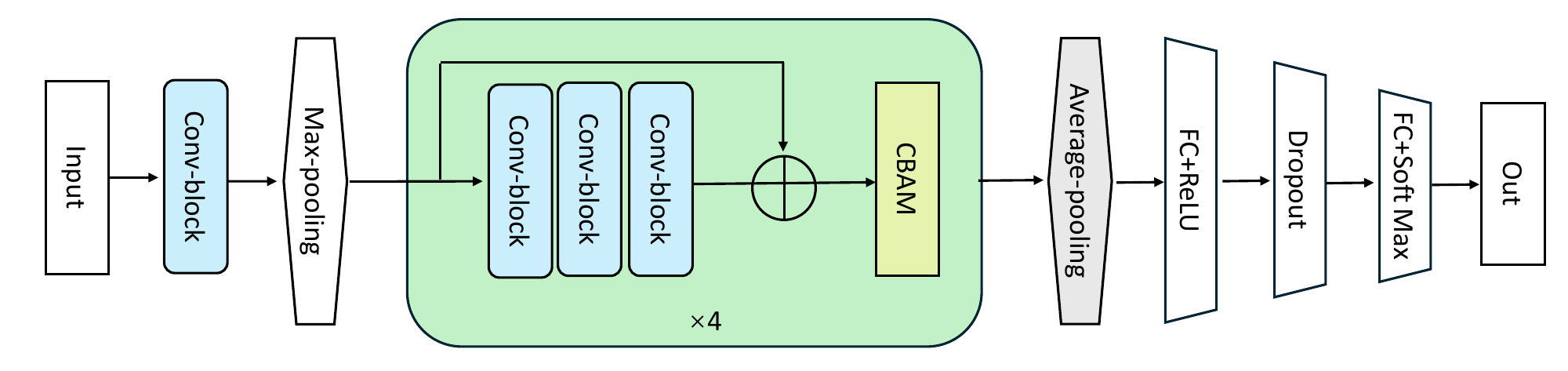}
\end{minipage}
\caption{Schematic diagram of our ResNet-CBAM architectures. The Conv-block includes a convolutional layer, an instance normalization layer, and an activation function. \label{fig3}}
\end{figure}

In this paper, we employ a DL model that combines Residual Networks (ResNet, \cite{2016cvpr.confE...1H}) and Convolutional Block Attention Module (CBAM, \cite{2018arXiv180706521W}) to classify GRBs. This model aims to enhance the capture of key features by introducing an attention mechanism, thereby improving classification performance. Figure \ref{fig3} shows a schematic diagram of our CNN architecture.

The input layer of the model accepts GRB count map data with a shape of $512\times 128\times 1$. First, the input data passes through a convolution block layer and an instance normalization layer \citep{2016arXiv160708022U}, using 64 $7\times7$ convolutional kernels with a stride of 2. This layer primarily extracts initial features from the input image and standardizes the data through the instance normalization layer to promote model convergence. Next, a $3\times3$ max-pooling layer further reduces the size of the feature map, thereby decreasing computational load and enhancing feature robustness.

The model integrates multiple residual blocks, each using shortcut connections with convolutions. Each residual block contains three Conv-blocks, each with 128 $3\times3$ convolutional kernels. After the last convolutional layer, a shortcut connection adds the input to form the residual. This structure mitigates the gradient vanishing problem in deep neural network training, improving training efficiency and accuracy. Each convolutional layer in the residual block is accompanied by instance normalization and Rectified Linear Unit (ReLU, \cite{10.5555/3104322.3104425}) activation functions to ensure stable training and non-linear expressiveness. Instance normalization helps reduce internal covariate shift, while ReLU activation introduces non-linearity, enhancing the model's expressiveness.

After each residual block, we introduce the CBAM to enhance the model's focus on important features. CBAM consists of two submodules: Channel Attention Module and Spatial Attention Module. The Channel Attention Module first applies global average pooling and global max pooling to the input feature map, generating two different feature descriptions. These features are then processed through two fully connected layers and merged to generate a channel weight vector. This weight vector is used to weight each channel of the input feature map, highlighting the features of important channels. The Spatial Attention Module generates spatial feature descriptions by computing the channel average and channel maximum of the feature map. After merging these descriptions, a convolutional layer generates the spatial attention map. This attention map weights each spatial position of the input feature map, highlighting important spatial features.

After processing through multiple residual blocks and CBAM modules, the feature map passes through a global average pooling layer and flattens into a one-dimensional vector. This vector then passes through a fully connected (FC) layer and ReLU activation function for feature extraction and dimensionality reduction. The fully connected layer contains 8 neurons and uses L1 and L2 regularization to prevent overfitting. To further prevent overfitting, a Dropout layer with a rate of 0.5 is added after the fully connected layer. This means that 50\% of the neurons are randomly dropped during each training process, reducing neuron co-adaptation and improving the model's generalization capability. Finally, a fully connected layer with a Softmax activation function outputs the classification results. The Softmax activation function is defined as:

\begin{equation}
\text{Softmax}(z_i) = \frac{e^{z_i}}{\sum_{c=1}^C e^{z_c}},
\end{equation}
where $z_i$ denotes the output of the classifier,  
i represents the category index.  C is total number of categories. This layer contains 2 neurons, used to output the probability of each classification category.

\subsection{Training and Optimization \label{Train}}
Our model is implemented using Keras \footnote{\url{https://keras.io}}  with TensorFlow \footnote{\url{https://www.tensorflow.org}} as the backend. The model is compiled using the Adam optimizer, an optimization algorithm that adapts learning rates based on estimates of first and second moments \citep{2014arXiv1412.6980K}. For this training, the initial learning rate is set to 0.0001, with $\beta 1$ and $\beta 2$ set to 0.95 and 0.999, respectively, and $\epsilon$ set to $10^{-8}$. The $\beta 1$ and $\beta 2$ parameters control the exponential decay rates for the first and second moment estimates, respectively. A higher $\beta 1$ value results in smoother first moment estimates, while a higher $\beta 2$ value provides more stable second moment estimates. The epsilon parameter prevents division by zero errors. The loss function used is the categorical cross-entropy loss function, suitable for multi-class classification problems. This loss function, defined as:
\begin{equation}
\text{Loss} = -\sum_{i=1}^N y_i \log(p_i),
\end{equation}
where $y_i$ is the true label and $p_i$ is the predicted probability, optimizes the model parameters to minimize the loss by calculating the cross-entropy between the predicted probability distribution and the true labels. The evaluation metric chosen is accuracy, used to assess the model's classification performance on the validation set.

After compiling the model, we define a custom callback function to record and print the maximum loss and accuracy on the validation set at the end of each epoch. This callback initializes the maximum loss and accuracy to 0 at the start of training and updates these values if the current validation accuracy exceeds the recorded maximum. It then prints the updated information. This callback helps monitor the model's performance during training, making it easier to observe how the model performs on the validation set.

Additionally, several callback functions are used during model training to optimize the training process. First, the \texttt{keras.callback.ReduceLROnPlateau} callback monitors the model's loss during training and reduces the learning rate when the loss plateaus. The factor parameter is set to 0.5, meaning the learning rate is multiplied by 0.5 each time it is reduced; the patience parameter is set to 20, meaning the learning rate is reduced after 20 epochs if the loss does not decrease; the minimum learning rate is set to 0.00005 to prevent it from becoming too low to effectively update the model parameters. Second, the \texttt{keras.callback.EarlyStopping} callback monitors the validation accuracy and stops training early if the accuracy does not improve. The patience parameter is set to 40, meaning training stops after 40 epochs if the accuracy does not improve. Third, the \texttt{keras.callback.ModelCheckpoint} callback saves the model with the best performance on the validation set, with the filename including the epoch, validation accuracy, and validation loss for easy reference. This callback monitors the validation accuracy and saves the model whenever a new highest accuracy is reached. Finally, the \texttt{keras.callback.CSVLogger} callback records log information during training, including training and validation loss, and accuracy for each epoch, saving the log file in CSV format for subsequent analysis and visualization.

Ultimately, the model is trained on the training and validation sets. Training parameters include 100 epochs, a batch size of 64, and detailed log output settings. During training, the model automatically adjusts the learning rate, saves the best model, and logs the training process. These steps allow the model to continuously optimize its parameters, improve validation accuracy, and eventually produce a well-performing classification model. This training process is conducted in an environment equipped with an NVIDIA GTX-3060 GPU.

\subsection{Feature Extraction and Visualization \label{Feature}}
By combining ResNet and CBAM, our model extracts more detailed and significant features at multiple levels. We can observe the characteristics of each GRB through the features extracted from each GRB count map by the model. In this work, we primarily study the feature maps from the average pooling layer. This layer performs average pooling on the input feature maps to reduce their size while retaining the main feature information. Feature maps are high-dimensional and complex datasets, making it challenging to distinguish and compare GRBs. By reducing the high-dimensional data and representing it in 2D or 3D, we can directly visualize patterns in the data distribution.

\cite{JMLR:v9:vandermaaten08a} proposed a nonlinear dimensionality reduction algorithm, t-distributed Stochastic Neighbor Embedding (t-SNE), which is highly valuable in large data analysis and is widely used in astrophysical data processing. Here, we apply the t-SNE algorithm to analyze the extracted feature maps. We implement t-SNE using \texttt{sklearn.manifold.TSNE \footnote{\url{https://scikit-learn.org/stable/modules/generated/sklearn.manifold.TSNE.html}}}.

When using t-SNE, the choice of several key parameters significantly impacts the quality and interpretability of the results. First, the \texttt{n\_components} parameter determines the target dimensionality after reduction, typically set to 2 or 3 for easy visualization. Second, the \texttt{perplexity} parameter controls the number of effective neighbors for each point, influencing the representation of local and global data structures. Lower \texttt{perplexity} values help reflect the local structure of the data, while higher values reveal more global structures. Typically, the value of \texttt{perplexity} is adjusted between 5 and 50. Choosing an appropriate value based on the dataset's characteristics can significantly improve visualization. In this work, we determine the optimal perplexity value to be 20 through grid search. The \texttt{n\_iter} parameter determines the number of iterations for gradient descent. Higher iterations are necessary to ensure the stability of the results. In our experiments, we set \texttt{n\_iter} to 15,000 to ensure stable t-SNE dimensionality reduction results. Other parameters use default settings.

\section{Result \label{Result}}
\subsection{Model Performance}
The model is trained according to the aforementioned training strategy. Figure \ref{fig2} shows the variations in loss and accuracy during the training and validation processes. The figure shows that the loss value decreases rapidly at the beginning of training, indicating that the model learns effective features during the initial phase. As the number of training epochs increases, the loss value stabilizes, and the loss values on the training and validation sets remain close, suggesting that the model does not exhibit significant overfitting during this training process. Throughout the training, the loss gradually decreases, and accuracy continuously increases. The consistent performance on both the training and validation sets shows the absence of significant overfitting, demonstrating good training effectiveness.

We further evaluate the best-performing classification model using the test set and present its performance through four metrics: Accuracy, Precision, Recall, and F1-score. Detailed descriptions of these parameters are provided in Appendix \ref{A}. Table \ref{tab1} presents the values of these four metrics for the three datasets. In the test set, the Accuracy is 99.40\%, Precision is 99.63\%, Recall is 99.63\%, and F1-score is 99.63\%.

The specific recognition rate for each class is represented by the confusion matrix. As shown in the test set confusion matrix (Figure \ref{fig5}), the model demonstrates excellent performance. For the ``GRB-S" class, 181 GRBs are correctly predicted, accounting for 98.37\% of the true ``GRB-S", while only 3 GRBs are misclassified. For the ``GRB-L" class, 814 bursts are correctly predicted, accounting for 99.63\% of the true ``GRB-L", while 3 GRBs are misclassified. These results indicate that the model has high accuracy in classifying the ``GRB-S" and ``GRB-L" categories, with very low error rates, further demonstrating the model's excellent performance and robustness.

\subsection{Application Result}
\subsubsection{Classification of overlapping GRBs}
We apply the optimal model to classify GRBs with overlapping $T_\text{90}$. As described in Section \ref{data}, we obtain 594 GRB events with overlapping $T_\text{90}$ and expand the sample to 2394 by considering data from four brighter detectors. After inputting these count maps into the optimal model, we can determine whether these GRB events are GRB-L or GRB-S.

Each GRB event corresponds to four detectors, and the number of photons detected by each detector may vary. This might lead to inconsistent classification results among the four detectors. To ensure an accurate classification, we use a voting mechanism to decide the final classification. For example, if three out of four detectors classify an event as GRB-L, we categorize the event as GRB-L. If the votes for GRB-L and GRB-S are equal, we consider the classification as a failure.

Out of all events, 562 GRBs are successfully classified, while 32 GRBs fail to be classified. The classification results are detailed in Table \ref{tab2}. Further analysis shows that the primary reason for classification failures is the insufficient photon counts for these GRBs. This affects the accuracy of the classification. As shown in Figure \ref{fig6}, the violin plot illustrates the distribution of durations for the two GRB categories. The $T_\text{90}$ distribution for GRB-S is more concentrated, with a maximum value of 8.192 s. The $T_\text{90}$ distribution for GRB-L is more dispersed, with a minimum value of 1.792 s.

We also extract and visualize the count map features of the overlapping GRBs, as described in Section \ref{Feature}. As shown in Figure \ref{fig7}, we present the 2D results of the dimensionally reduced feature maps. The figure shows that the overlapping GRBs are clearly divided into two parts. This indicates that our method effectively classifies the overlapping GRBs into two categories, indirectly disproving the existence of a third intermediate GRB type.

\subsubsection{Feature Extraction of GRB Count Maps}
We apply the optimal model to feature extraction of GRB count maps. We select the count map from the brightest detector for each GRB (a total of 3726) and extract features and visualize them using the method described in Section \ref{Feature}. We successfully extract features for 3726  GRBs. As shown in Figure \ref{fig8}, the 2D and 3D results of the dimensionally reduced feature maps are presented. The t-SNE mappings are colored based on $T_\text{90}$. Interestingly, the features of GRBs form two distinct groups after dimensionality reduction. The smaller cluster mainly consists of short-duration GRBs, while the larger cluster consists of long-duration GRBs. The distribution of $T_\text{90}$ values on the t-SNE mapping gradually increases in a certain direction, with colors changing from deep blue to red. This indicates that the features extracted by the optimal model effectively capture the temporal characteristics of GRBs.

To differentiate from traditional classification methods, this paper refers to the short duration cluster as S-type GRBs and the longer-duration cluster as L-type GRBs. There are 756 S-type GRBs, accounting for 20.3\% of the total GRBs, and 2970 L-type GRBs, accounting for 79.7\% of the total bursts. The statistical results of the two types of bursts are shown in Table \ref{tab3}. There is no absolute boundary between the $T_\text{90}$ of the two types of GRBs; some GRBs in the L-type cluster have shorter $T_\text{90}$ than some GRBs in the S-type cluster. The $T_\text{90}$ of GRB-S can be as long as 8 s, while the $T_\text{90}$ of GRB-L can be as short as 0.4 s, which differs from the traditional classification method that uses $T_\text{90} = 2$ s as the boundary.

\section{Discussion} \label{Discussion}
\subsection{GRBs Associated with Kilonovae and Supernovae}
After feature extraction and dimensionality reduction of the GRB count maps, the results show that bursts with similar temporal characteristics cluster closely together. These two clusters strongly suggest fundamentally different physical properties and/or origins. According to the established GRB origin theories, GRBs associated with supernovae are generally thought to originate from the collapse of massive stars, while GRBs associated with kilonovae are thought to originate from the merger of compact star binaries. Therefore, it is necessary to search for GRBs with additional electromagnetic burst signals in the Fermi sample and study their distribution in the t-SNE mapping to further explore the physical nature behind this classification. This will help verify whether our proposed classification method is consistent with existing observational results and reveal the physical mechanisms of different types of GRBs.

In the Fermi catalog, our kilonova sample includes five events: GRB 150101B \citep{2018NatCo...9.4089T,2019MNRAS.489.2104T}, 170817A \citep{2017Natur.551...71T}, 211211A \citep{2022Natur.612..232Y,2022Natur.612..228T,2022Natur.612..223R,2023NatAs...7...67G}, and 230307A \citep{2024ApJ...962L..27D,2024Natur.626..737L,2024Natur.626..742Y}. GRB 170817A is associated with GW170817 and has been confirmed to originate from the merger of neutron star binaries. GRB 211211A and GRB 230307A are different from other GRBs associated with kilonovae, as their durations far exceed 2 s. In Figure \ref{fig9}, we locate the positions of the GRBs associated with KN. In our classification results, GRB 150101B, 160821B, and 170817A are classified as S-type GRBs, while GRB 211211A and 230307A are classified as L-type GRBs. This indicates that GRB 211211A and 230307A are indeed very special and warrant further study of their observational properties and physical origins.

In fact, \cite{2022Natur.612..232Y} suggested that GRB 211211A may originate from a white dwarf-neutron star merger. Large-scale computer simulations by \citep{2023ApJ...954L..21G} studied the evolution of relativistic jets in the merger of black holes and neutron stars. The simulation results suggest that weaker magnetic fields can produce longer-duration jets, consistent with the observed characteristics of GRB 211211A. \citep{2023ApJ...947...55B} proposed that collapsars could also explain the origin of GRB 211211A. Similarly, independent studies of the temporal and spectral characteristics of these two GRBs by several authors also suggest that they may have the same origin. Thus, the physical origin of GRB 211211A and GRB 230307A requires further exploration and confirmation through more observations.

We also studied the distribution of GRBs associated with supernovae in the t-SNE mapping. Our supernova sample includes: GRB 091127A \citep{2010GCN.10400....1C,2011ApJ...743..204B}, GRB 101219B \citep{2011ApJ...735L..24S}, GRB 130215A \citep{2014A&A...568A..19C}, GRB 130427A \citep{2013ApJ...776...98X}, GRB 130702A \citep{2013GCN.14998....1C,2013ApJ...776L..34S}, GRB 140606B \citep{2015MNRAS.452.1535C}, GRB 171010A \citep{2019MNRAS.490.5366M}, GRB 180728A \citep{2019ApJ...874...39W}, GRB 190114C \citep{2022A&A...659A..39M}, GRB 200826Ac\citep{2022ApJ...932....1R}, GRB 211023A \citep{2022GCN.31596....1R,2023ApJ...955...93A}, and GRB 230812B \citep{2024ApJ...960L..18S}. The distribution of these bursts in our classification results is shown in Figure \ref{fig9}. Except for GRB 200826A, the remaining 12 bursts associated with supernovae are all classified as L-type GRBs.

GRB 200826A is a remarkable GRB whose characteristics challenge traditional classification standards. Studies have found that GRB 200826A was produced by a nascent black hole formed from the collapse of a massive star \citep{2021NatAs...5..917A,2021NatAs...5..911Z}. Additionally, observations of GRB 200826A show optical afterglows similar to those of supernova explosions, further confirming its origin from the collapse of a massive star. This finding brings a new perspective to GRB classification, indicating that some short bursts may actually be produced by collapsars.

Recently, \cite{2024ApJ...963L..12P} proposed new insights into the origins of GRBs. They analyzed the formation rates of SGRBs and LGRBs, finding that the formation rate of low-redshift LGRBs is significantly higher than the cosmic star formation rate and similar to that of SGRBs, which is inconsistent with traditional models. Their results predict that about 60\% ± 5\% of LGRBs with redshifts less than 2 have formation rates similar to those of SGRBs and decline rapidly with increasing redshift. This aligns with the expectation of delayed compact star mergers, suggesting that this portion of LGRBs likely originates from compact star mergers. The remaining 40\% ± 5\% of LGRBs follow the cosmic star formation rate and may originate from the collapse of massive stars. These LGRBs may be associated with supernova explosions. Some observational evidence supports the hypothesis that compact star mergers are the origin of a subset of low-redshift LGRBs. For instance, the recent discovery of two low-redshift LGRBs (GRB 211211A and GRB 230307A) is associated with a kilonova. Kilonovae are believed to be products of compact star mergers, providing strong support for the possibility that some low-redshift LGRBs originate from compact star mergers.

\subsection{Short GRBs with Extended Emission}

In SGRBs, a subset is characterized by a short/hard spike, followed by a series of longer-lasting soft  pulses. These GRBs are known as short GRBs with extended emission (sGRB-EE) \citep{2006ApJ...643..266N}. GRB 060614 is a long event lasting over 100 s, with a light curve initially showing a short pulse followed by extended emission \citep{2006Natur.444.1050D,2006Natur.444.1053G,2007A&A...474L..13B,2008AIPC.1065...39L} This burst did not detect a supernova and had delays and luminosity consistent with short GRBs, suggesting a possible merger origin \citep{2006Natur.444.1050D}. Since the first clear evidence of extended emission was found in GRB 060614, researchers have extensively searched and studied more such events in the Swift and Fermi satellite eras \citep{2010ApJ...717..411N,2015MNRAS.452..824K,2020MNRAS.492.3622L}.

In this section, we investigate the distribution of sGRB-EE in our classification results. Our sample, taken from the work of \cite{2020MNRAS.492.3622L} and \cite{2023ApJ...954L...5V}, includes 38 sGRB-EE, with 35 having $T_\text{90} > 2$ s and 3 having $T_\text{90} < 2$ s. The SGRB-EE samples are given in Table \ref{tab3}. Figure \ref{fig10} shows the sGRB-EE marked in the new classification results. From the figure, it can be seen that most sGRB-EE with $T_\text{90} > 2$ s are located in the L-type GRB region, with only one exception, indicating that sGRB-EE characteristics are consistent with L-type GRBs. Conversely, sGRB-EE with $T_\text{90} < 2$ s are located in the S-type GRB region, indicating that these bursts share characteristics with S-type GRBs.

For sGRB-EE, we find that most are located in the L-type GRB region, with a few in the S-type GRB region. This suggests that sGRB-EE in our classification method can exhibit features of either LGRB or SGRB, depending on their duration. This finding further validates the effectiveness of our classification method and demonstrates its advantages in handling complex GRB events.

\subsection{Comparative analysis of spectral characteristics in two types of GRBs}
Spectral analysis is an important method for exploring the physical properties of GRBs. Spectral parameters such as hardness ratio (HR), peak energy ($E_\text{p}$), and spectral index ($\alpha$) are commonly used for GRB classification. To study the physical properties of S-type and L-type GRBs, we performed time-integrated spectral analysis on all GRBs in the sample, with detailed analysis procedures provided in Appendix \ref{B}. In this section, we compare the spectral properties of the two types of GRBs.

As shown in Figure \ref{fig11}, the distribution of the $\alpha$ in L-type and S-type GRBs is illustrated. In Figure \ref{fig11}(a), each point represents an individual GRB event, with the color indicating the value of $\alpha$. The color bar ranges from -1.0 (blue) to 0.5 (red), showing the distribution of $\alpha$ values across events. Although some data points have similar colors, the overall clustering structure remains clear. In Figure \ref{fig11}(b), the violin plot combines a box plot and a density plot to show the distribution shape and concentration trend of $\alpha$. The $\alpha$ distribution of L-type GRBs is wider, while that of S-type GRBs is relatively concentrated. The black box in the plot indicates the interquartile range, and the white dot represents the median. It can be observed that the median of S-type GRBs is slightly higher than that of L-type GRBs, while the interquartile range is roughly the same for both types, indicating that S-type GRBs tend to have a slightly higher $\alpha$ distribution than L-type GRBs.

Figure \ref{fig12} shows the distribution of $E_\text{p}$ for L-type and S-type GRBs. Among L-type GRBs, the number of GRB events is higher, and the energy range is wider, mainly concentrated in the lower energy region (mostly blue), but there are also some high-energy events (red points). This indicates that L-type GRBs have a diverse energy distribution, ranging from low to high values. For S-type GRBs, the number of GRB events is relatively smaller, but the energy distribution also covers a wide range from low to high. Although the total number of events is fewer, they still show a diverse energy distribution. Additionally, in the violin plot (Figure \ref{fig12}(b)), it can be seen that the peak energy distribution of L-type GRBs is wider, indicating a larger energy span for this group. The median position is slightly lower than that of the S group, suggesting that the L group has more low-energy events, but there are also some high-energy events. The overall distribution shape shows that the data are more dispersed. In S-type GRBs, the peak energy distribution is narrower, indicating that the energy is concentrated within a smaller range. The median of the S group is slightly higher than that of the L group, indicating that S-type GRBs have relatively higher peak energies, but the range is smaller, and the data are more concentrated. These two figures show that the energy distribution characteristics of L-type and S-type GRBs are significantly different. L-type GRBs have a broader energy distribution, including more low-energy events and some high-energy events, with more dispersed data, possibly related to their longer durations and complex radiation mechanisms. S-type GRBs have a relatively concentrated energy distribution, with higher peak energies but a smaller range, indicating more concentrated data, possibly related to their shorter durations and different radiation mechanisms.

Figure \ref{fig13} shows the distribution of Flux for L-type and S-type GRBs. Similar to $E_\text{p}$, the flux distribution of L-type GRBs is broader, including both high and low flux events. The median is slightly lower than that of S-type GRBs. The flux distribution of S-type GRBs is narrower, concentrated within a smaller range, and the median is slightly higher than that of L-type GRBs. Overall, the flux distribution of L-type GRBs is more dispersed, while that of S-type GRBs is more concentrated.

In addition to the above spectral parameters, we also calculated the HR for both groups of GRBs, as detailed in Appendix \ref{B}. Some studies suggest that there may be a correlation between the hardness of GRB spectra and their classification \citep{1993ApJ...413L.101K,2019ApJ...870..105T}. Although hardness alone may not clearly separate SGRBs and LGRBs, there may still be a strong correlation between hardness and type, particularly since the hardness of a burst should be closely related to its physical origin. As shown in Figure \ref{fig14}, the HR distribution of L-type and S-type GRBs is illustrated. In the t-SNE mapping (Figure \ref{fig14}(a), harder GRBs cluster together and are all classified as S-type. However, hard GRBs are present in both S-type and L-type groups. Overall, most S-type GRBs are harder than most L-type GRBs, but hardness alone is not sufficient to determine whether a GRB is S-type or L-type. We note that the HR distribution is similar to the $E_\text{p}$ distribution.

Furthermore, we applied the two-dimensional Kolmogorov-Smirnov (KS) test to verify the statistically significant differences in $\alpha$, HR, $E_\text{p}$, and Flux between L-type and S-type GRBs. Table \ref{tab5} shows the results of the KS test. The KS statistic displayed in the table is the maximum difference between the two cumulative distribution functions. The p-value is used to test the significance of the null hypothesis (that the two groups of data come from the same distribution). Since all p-values are below 1.0\%, this hypothesis is rejected \citep{lazzeroni2014p,boos2011p}. This indicates that the differences in these spectral parameters between L-type and S-type GRBs are significant.

\subsection{Comparison with traditional classification methods}
In this study, we identified two distinct clusters corresponding to S-type and L-type GRBs through feature extraction and dimensionality reduction analysis of GRB count maps. We compare these two classification results with traditional classification methods.

The Amati relation is based on the energy and spectral characteristics of GRBs, specifically the relationship between peak energy ($E_\text{pi}$) and isotropic equivalent energy ($E_\text{iso}$)\citep{2002A&A...390...81A}. According to the Amati relation, LGRBs usually follow a positive correlation between $E_\text{pi}$ and $E_\text{iso}$, while SGRBs often deviate from this relationship. We analyzed GRBs with existing redshift data in the Fermi catalog (a total of 185), derived $E_\text{pi}$ and $E_\text{iso}$ values, and provided detailed calculations in Appendix \ref{C}. As shown in Figure \ref{fig16}, the relationship between $E_\text{pi}$ and $E_\text{iso}$ for GRBs with $T_\text{90} < 2$ s and $T_\text{90} > 2$ s is presented, showing that the two types of bursts follow different trajectories. Additionally, some special GRB events, such as GRB 200826A, GRB 230307A, GRB 211211A, and GRB 170817A, are marked in the figure. Notably, GRB 211211A and GRB 230307A have durations much longer than 2 s but appear on the short burst trajectory; conversely, GRB 200826A has a duration less than 2 s but falls on the long burst trajectory. There is no clear boundary between long and short bursts in the Amati relation, with significant overlap.

Based on this, \cite{2020MNRAS.492.1919M} proposed a new GRB classification method. They defined a new parameter, EHD (Energy-Hardness-Duration), which combines $E_\text{p,i}$ and $E_\text{iso}$ parameters to differentiate GRB types. Additionally, they included burst duration ($T_\text{90,i}$) in the classification based on EHD. Details of these two parameters are provided in Appendix \ref{C}.

As shown in Figure \ref{fig15}(a), the relationship between  $\rm log10(T_{90,i})$ and $\rm log_{10}(EHD)$ is illustrated, with SGRBs ($T_\text{90}<2$ s) and LGRBs ($T_\text{90}>2$ s) distinguished by color and markers. The black dashed line serves as the classification boundary, clearly separating GRBs into Type I and Type II, with the upper left region mainly containing SGRBs and the lower right region mainly containing LGRBs. This boundary provides an effective method for distinguishing the two GRB types, supporting the validity of using the EHD parameter for GRB classification. We compared the results of the EHD classification method with those of the new classification method, as shown in Figure \ref{fig15}(b). The figure shows that most GRBs are classified consistently, but a few bursts are classified as Type II in the EHD method and as S-type in our classification.

We specifically focus on some marked special GRBs that were discussed in the previous section. GRB 170817A is associated with a binary neutron star merger and the gravitational wave event GW170817. Under the EHD classification method, GRB 170817A is labeled as a Type I GRB, while in the new classification method, it is labeled as an S-type GRB. GRB 200826A has a duration of less than 2 s; under the EHD classification method, it is shown as a Type I GRB, while our new method classifies it as an S-type GRB. For the special events GRB 230307A and GRB 211211A, both the EHD classification method and the new method classify them as Type II (L-type) GRBs.

In addition to the Amati relation, the Dainotti relation is also one of the important empirical relations in GRB research. This relation describes the anticorrelation between the end time of the plateau phase in X-ray afterglows ($T_\text{a}$) and the corresponding X-ray luminosity ($L_\text{X}$) \citep{2008MNRAS.391L..79D,2011ApJ...730..135D,2010ApJ...722L.215D,2017A&A...600A..98D}. Subsequently, by introducing the prompt emission peak luminosity ($L_\text{peak}$) into the two-dimensional (2D) relation, a three-dimensional (3D) relation was obtained that describes the correlation among $L_\text{X}$, $T_\text{a}$, and $L_\text{peak}$ (also called ``fundamental plane relation", \cite{2016ApJ...825L..20D,2020ApJ...903...18S}). As research deepens, the Dainotti relation has been extended to optical and radio afterglows \citep{2020ApJ...905L..26D,2022ApJS..261...25D}, and is commonly observed in different types of GRBs, including short GRBs, GRBs associated with supernovae, and X-ray flashes \citep{2020ApJ...903...18S}. \cite{2016ApJ...825L..20D} separated GRBs into a subclass called ``Gold" sample based on plateau phase characteristics and defined a tight 3D fundamental plane relation, where the ``Platinum" sample is a subset of the Gold sample \citep{2022MNRAS.512..439C}. Compared to the Amati relation, the platinum sample lies on the long GRB trajectory and within the $1 \sigma$ confidence interval (see Figure \ref{fig16}); in our classification results, all samples are L-type GRBs (see Figure \ref{fig15}). The Dainotti relation reveals the connection between GRB afterglows and instantaneous radiation, providing important clues for understanding its physical mechanisms and classifications. However, the number of GRBs in the Platinum sample that are consistent with ours is small, making more detailed comparisons impossible. As observational data accumulates, future analyses of larger samples of GRBs may explore the connection between the Dainotti relation and other empirical relations in GRBs, as well as the differences among different types of GRBs.

\subsection{ Comparison with Other Machine Learning Classification Results}
\cite{2020ApJ...896L..20J} applied unsupervised machine learning algorithms, t-SNE and UMAP, to analyze GRB light curves and successfully classified 2294 Fermi GRBs into short and long bursts. However, their study found that a small subset of GRBs could not be robustly classified. We compare our classification method with their results. In our sample, 2268 GRBs overlapped with theirs. The results show that the two methods agreed on the classification of 2121 GRBs, accounting for 93.5\% of the total sample. Notably, in their method, 43 GRBs could not be robustly classified, while in our method, all these GRBs were successfully classified.

Additionally, we compare the classification results for overlapping GRBs. In their sample, 344 GRBs were classified consistently with our sample, of which 21 were considered unclassifiable by their method. In contrast, our method agreed with theirs on 196 GRBs, accounting for 60.9\% of the total sample. Among the 21 unclassifiable GRBs, our method successfully classified 20 as either L-type or S-type GRBs. However, GRB 110916016 was not successfully classified by our method either.

Apart from classifying GRBs based on light curves, \cite{2024MNRAS.527.4272C} applied t-SNE and UMAP to two samples of time-resolved and time-integrated spectral parameters from Fermi GRBs, successfully classifying them into two categories. We compare their classification results with ours, where 2271 GRBs overlapped with our sample. The results show that, in their time-resolved spectral sample, 2161 GRBs were classified consistently with our results, accounting for 95.2\% of the total sample; in the time-integrated spectral sample, 2149 GRBs (94.6\% of the total sample) were classified consistently with our results.

Moreover, \cite{2019Ap&SS.364..105H} used principal component analysis and statistical clustering techniques to study a sample of 801 GRBs described by 16 variables and identified three optimal GRB categories. The first category aligns with the well-known short GRB category, while the other two are long GRB categories. In contrast, our classification method rejects the existence of a third intermediate GRB category, a result that aligns with the classification results obtained by \cite{2023ApJ...959...44L} using supervised learning methods.

\cite{2023MNRAS.525.5204B} utilized unsupervised machine learning methods to analyze optical and X-ray samples of GRBs with plateau phases, revealing some significant ``microtrends.” In the optical sample, X-ray-rich (XRR), X-ray flash (XRF), and GRBs accompanied by supernovae (GRB-SNe) generally cluster with LGRB. However, ultra-long GRBs (ULGRB, $T_\text{90} > 1000$ s) do not cluster with LGRB, suggesting they may have different origins. In the X-ray sample, XRR and GRB-SNe similarly cluster with LGRB, while XRF also shows a similar trend, though it differs under certain parameter combinations. SGRB, intrinsically short GRBs (IS, $\rm T_\text{90}/(1 + z) < 2$ s), and short extended emission GRBs (SEE) show more complex clustering results; they sometimes cluster with LGRB but at other times separate independently depending on parameter combinations. We compared these results with current classification methods using sample data from \cite{2022ApJ...940..169D} and \cite{2023ApJS..267...42D}. Figure \ref{fig17} shows the distribution of IS, UL, XRR, and XRF in our classification. The analysis suggests that XRR and XRF are classified as L-type GRBs in our system, consistent with the clustering results by \cite{2023MNRAS.525.5204B}, indicating that these GRB types may have similar origins or share a common origin but in different environments. Of the four IS samples, three are robustly classified as S-type GRBs, while GRB 090423 is classified as an L-type GRB. Studies suggest that GRB 090423 may originate from the collapse of a massive star, as discussed by \cite{2009Natur.461.1258S}. For ULGRBs, we robustly classify them as L-type GRBs, though this result contradicts the analysis by \cite{2023MNRAS.525.5204B}. The small sample size of ULGRBs may affect the stability of the clustering analysis.

\section{Summary and Conclusion \label{Summary}}
In this paper, we propose a new classification scheme based on DL for precise classification of GRBs, specifically addressing the challenge of overlapping GRB classifications in $T_\text{90}$. First, we use the BGMM model to fit $T_\text{90}$ data from 3774 GRBs and identify overlapping GRB events. By setting a posterior probability threshold, we confirm 594 overlapping GRBs as classification samples. To further expand the training sample, we utilize the multi-detector data from Fermi-GBM, treating each detector as a separate GRB. This approach generated a large-scale dataset, providing a rich and high-quality data foundation for training DL models.

For model construction, we employ a DL model combining ResNet and CBAM. This model enhances the capture of image features by introducing attention mechanisms, significantly improving classification performance. The model's input consists of processed GRB count maps (containing light curves and spectral information). Through multiple convolutional, normalization, pooling, and residual blocks, the model extracts complex and crucial features. Each residual block is followed by a CBAM module, which strengthens the model's focus on important features through channel and spatial feature weighting. Finally, the model outputs the probability distribution of GRB classifications via a fully connected layer and Softmax activation function. We conduct a detailed evaluation of the model's performance during training and validation. The model was trained using the Adam optimizer with dynamically adjusted learning rates to prevent overfitting. The loss values on both the training and validation sets rapidly decreased and stabilized, indicating effectiveness. Ultimately, the optimal model achieved an accuracy of 99.40\%, precision of 99.63\%, recall of 99.63\%, and F1 score of 99.63\% on the test set, demonstrating excellent performance in GRB classification.

We apply the optimal model to classify $T_\text{90}$ overlapping GRBs with promising results. The results show that out of all $T_\text{90}$ overlapping events, 594 GRBs were successfully classified into two categories (GRB-S and GRB-L). GRB-S had a more concentrated $T_\text{90}$ distribution with a maximum value of 8.192 s, while GRB-L had a more dispersed distribution with a minimum value of 1.792 s.

Additionally, we use the trained model as a feature extractor to extract features from count maps of 3376 GRBs. Using t-SNE dimensionality reduction, we successfully map the high-dimensional feature space to a low-dimensional space, revealing two distinct clusters. One smaller cluster consisted mainly of GRBs with shorter durations, while the larger cluster comprised GRBs with longer durations, named S and L type GRBs, respectively. There is no absolute boundary between the $T_\text{90}$ values of the two GRB types; S-type GRBs can have $T_\text{90}$ up to 8 s, while L-type GRBs can be as short as 0.4 s, differing from the traditional classification method with $T_\text{90}=2$ s as the boundary. We analyze and discuss these two GRB types in detail, and the results are summarized as follows:

(1)Compared to $T_\text{90}$ based classification methods, the DL approach demonstrated higher effectiveness in GRB classification, particularly in handling overlapping $T_\text{90}$ GRBs. Our method also refute the existence of intermediate categories.
We analyze the distribution of GRBs related to KN and SN in the t-SNE map to explore the underlying physical properties. In KN samples, except for GRBs 211211A and 230307A, all others were classified as S-type GRBs. GRBs 211211A and 230307A showed unique characteristics, suggesting they might have atypical physical origins, warranting further research. For SN-related GRBs, except for GRB 200826A, all other events are classified as L-type GRBs. The characteristics of GRB 200826A challenge traditional classification standards, suggesting that some GRBs considered short bursts might actually originate from collapsing stars.

(2)In the classification of sGRB-EE, we find that sGRB-EE with $T_\text{90}>2$ s mostly fall into the L-type GRB region, while those with $T_\text{90}<2$ s fall into the S-type GRB region, indicating that the duration characteristics of sGRB-EE can be similar to either LGRB or SGRB.

(3)Through time-integrated spectral analysis of GRBs in the sample, we compare the spectral parameter distributions between the two GRB types and provide a spectral parameter catalog. The results show that S-type GRBs generally have higher low-energy spectral indices  and peak energies, with energy concentrated in higher ranges and higher flux. L-type GRBs exhibit a broader distribution of these parameters, including some high-energy events despite most having lower energy. Additionally, S-type GRBs usually have higher spectral hardness, but HR overlapped between the two GRB types. Overall, S-type GRBs show higher and more concentrated spectral parameters, while L-type GRBs have a broader parameter range. A two-dimensional KS test further confirm significant differences in spectral characteristics between the two GRB types.

(4)We compare the two GRB types obtained from feature extraction and dimensionality reduction of GRB count maps with traditional classification methods. The conventional Amati relation is commonly used to distinguish LGRB and SGRB but cannot fully differentiate them due to overlap. The EHD parameter classification method provides clearer standards but relies on redshift observations. In contrast, the new classification method can more accurately distinguish S-type and L-type GRBs without relying on redshift observations, showing significant advantages over traditional methods.

In summary, compared to traditional methods, the classification strategy proposed in this study demonstrates stronger robustness and effectiveness in identifying GRB types and can be applied to classify newly discovered GRBs in the future.

\section{Data Availability}
The code and data sets are available upon reasonable request.

\begin{acknowledgments}

We acknowledge the use of the public data from the Fermi data archives. This work is supported by the National Natural Science Foundation of China (NSFC 12233006 and 12163007), the Yunnan University graduate research innovation fund project KC-23233826, and Scientific Research Fund of Education Department of Yunnan Province (2024Y036).
\end{acknowledgments}

\vspace{5mm}

\software{GBM-Data-Tools \citep{GbmDataTools}, matplotlib \citep{2007CSE.....9...90H}, numpy \citep{2020Natur.585..357H}, pandas \citep{2022zndo...3509134T}, scipy \citep{2020NatMe..17..261V}, scikit-learn \citep{scikit-learn}, tensorflow \citep{tensorflow_developers_2024_12726004}, 3ML \cite{2015arXiv150708343V}.}

\bibliography{Manuscrip}{}
\bibliographystyle{aasjournal}

\appendix
\section{Appendix information \label{A}}
In the classification of GRBs, the performance is typically described by four metrics: accuracy, precision, recall, and F1 score, as follows:

Accuracy represents the proportion of correctly classified samples (including both short GRBs and long GRBs) among all samples. The formula is:
\begin{equation}
\text{Accuracy} = \frac{TP + TN}{TP + TN + FP + FN}.
\end{equation}
Where TP (True Positive) is the number of long GRBs correctly predicted as long GRBs, TN  (True Negative) is the number of short GRBs correctly predicted as short GRBs, FP (False Positive) is the number of short GRBs incorrectly predicted as long GRBs, and FN (False Negative) is the number of long GRBs incorrectly predicted as short GRBs. Accuracy provides an overall performance overview, but it might not be sufficient in cases of class imbalance.

Precision represents the proportion of actual long GRBs among the samples predicted as long GRBs. The formula is:
\begin{equation}
\text{Precision} = \frac{TP}{TP + FP}.
\end{equation}
Precision reflects the reliability of the model in predicting long GRBs, especially in scenarios where reducing false positives (short GRBs misclassified as long GRBs) is crucial.

Recall represents the proportion of correctly predicted long GRBs among all actual long GRBs. The formula is:
\begin{equation}
\text{Recall} = \frac{TP}{TP + FN}.
\end{equation}
Recall indicates the model's ability to detect long GRBs, particularly important in scenarios where reducing false negatives (long GRBs misclassified as short GRBs) is crucial.

The F1 score is the harmonic mean of precision and recall, used to comprehensively evaluate the model's performance in classifying long GRBs. The formula is:
\begin{equation}
\text{F1 Score} = 2 \times \frac{\text{Precision} \times \text{Recall}}{\text{Precision} + \text{Recall}}.
\end{equation}
The F1 score provides a balanced evaluation between precision and recall, suitable for scenarios with imbalanced classes of long GRBs and short GRBs.

\section{Spectral analysis \label{B}}
We used a Bayesian analysis package, specifically the Multi-Mission Maximum Likelihood framework (3ML;\cite{2015arXiv150708343V}), to perform time-integrated spectral analysis on the GRBs in our sample. This Bayesian method is widely applied in GRB spectral analysis within the Fermi-GBM catalog \citep{2019ApJ...886...20Y}. In our previous studies, we have successfully applied this method for data analysis (e.g., \cite{2021ApJ...920...53C,2024ApJ...964...45C}). During data analysis, we selected the three brightest NaI detectors and the brightest BGO detector, using energy ranges of NaI: 8–800 keV and BGO: 250 keV–30 MeV. The time interval for the time-integrated spectrum was determined by the start and end times of $T_\text{90}$, ensuring the inclusion of the entire prompt emission of the burst.

Since the radiation model of GRBs remains an unresolved issue and our study focuses on a comprehensive analysis of the spectral properties of the two types of GRBs, we chose to use only the Band function \citep{1993ApJ...413..281B} to examine the spectral characteristics of the time-integrated spectrum over the entire duration of the burst. The expression of the Band function is as follows:
\begin{equation}
N(E)_{\text{Band}} = A_{\text{BAND}} \times 
\begin{cases} 
\left(\frac{E}{100 \, \text{keV}}\right)^{\alpha} \exp\left[-\frac{E(2+\alpha)}{E_p}\right], & E \leq \frac{\alpha - \beta}{2 + \alpha} E_p \\[10pt]
\left(\frac{(\alpha - \beta) E_p}{(2 + \alpha)100 \, \text{keV}}\right)^{(\alpha - \beta)} \exp(\beta - \alpha) \left(\frac{E}{100 \, \text{keV}}\right)^{\beta}, & E \geq \frac{\alpha - \beta}{2 + \alpha} E_p,
\end{cases}
\end{equation}
where A is the normalization constant in the unit of $ph\ cm^{-2}kev^{-1}s^{-1}$. $\alpha$ and $\beta$ are the low-energy and high-energy  power-law spectral indices, respectively. $E_\text{p}$ is the peak energy in the $\nu F_{\nu}$ spectrum, unit in keV. We successfully performed time-integrated spectral analysis on 3,296 GRBs in our sample, with the results summarized in Table \ref{tab4}. In the table we show the main spectral parameters and their GRB classification results.

Based on the spectral analysis results, we calculate the hardness ratio (HR) for each GRB, as shown in Table B. The HR for each GRB is determined by the ratio of the fluence in the energy range 10-50 keV to the fluence in the energy range 50-300 keV, as defined below:
\begin{equation}
\text{HR} = \frac{\int_{50 \, \text{keV}}^{300 \, \text{keV}} E N(E)_\text{Band} dE}{\int_{10 \, \text{keV}}^{50 \, \text{keV}} E N(E)_\text{Band} dE}.
\end{equation}

\section{Amati relationship and EHD parameters \label{C}}
\cite{2002A&A...390...81A} identified a correlation between the energy $E_\text{p,i}$ and $E_{\text{iso}}$, which has since become the subject of numerous publications. This relationship requires redshift measurements to determine the intrinsic properties of the source. In our sample, redshift information is available for 259 GRBs. Based on the aforementioned spectral analysis results, we calculate:
\begin{equation}
E_{\text{p,i}} = E_{\text{p}}(1+z),
\end{equation}

\begin{equation}
E_{\text{iso}} = 4 \pi d_L^2 S_{\text{bol}}k (1 + z)^{-1},
\end{equation}
where $E_\text{p}$ is derived from the time-integrated spectral analysis. The bolometric fluence $S_{\text{bol}}$ is calculated from the observed radiation flux within the 1-1000 keV energy band in the rest frame. The correction factor k, as defined by \cite{2001astro.ph..2371B}, is given by:

\begin{equation}
k = \frac{\int_{1 \, \text{keV}/(1+z)}^{10^{4} \, \text{keV}/(1+z)} E N(E) dE}{\int_{8 \, \text{keV}}^{10^4 \, \text{keV}} E N(E) dE},
\end{equation}
The cosmological distance \( d_L \) is calculated using the following equation:
\begin{equation}
d_L = \frac{(1+z) c}{H_0} \int_{0}^{z} \frac{dz'}{\sqrt{\Omega_M (1+z')^3 + \Omega_L}}.
\end{equation}
We adopt the following cosmological parameters: $\Omega_M = 0.27$, $\Omega_L = 0.73$, and $H_0 = 70\ km\ s^{-1}\ Mpc^{-1}$. As shown in Figure \ref{fig16}, we present the relationship between $E_\text{{iso}}$ and $E_\text{p,i}$.

The Energy-Hardness-Duration (EHD) parameter is calculated based on the values of $E_{p,i}$, $E_{\text{iso}}$, and $T_\text{90}$ as follows \citep{2020MNRAS.492.1919M}:

\begin{equation}
EHD = \frac{(E_{p,i}/100 \, \text{keV})}{(E_{\text{iso}}/10^{51} \, \text{erg})^{0.4} \, (T_{90,i}/1 \, \text{s})^{0.5}}.
\end{equation}
where $T_\text{90,i} = T_\text{90}/(1+z)$.

\begin{figure}[htbp]
\centering
\begin{minipage}[t]{1.0\textwidth}
\centering
\includegraphics[width=\textwidth]{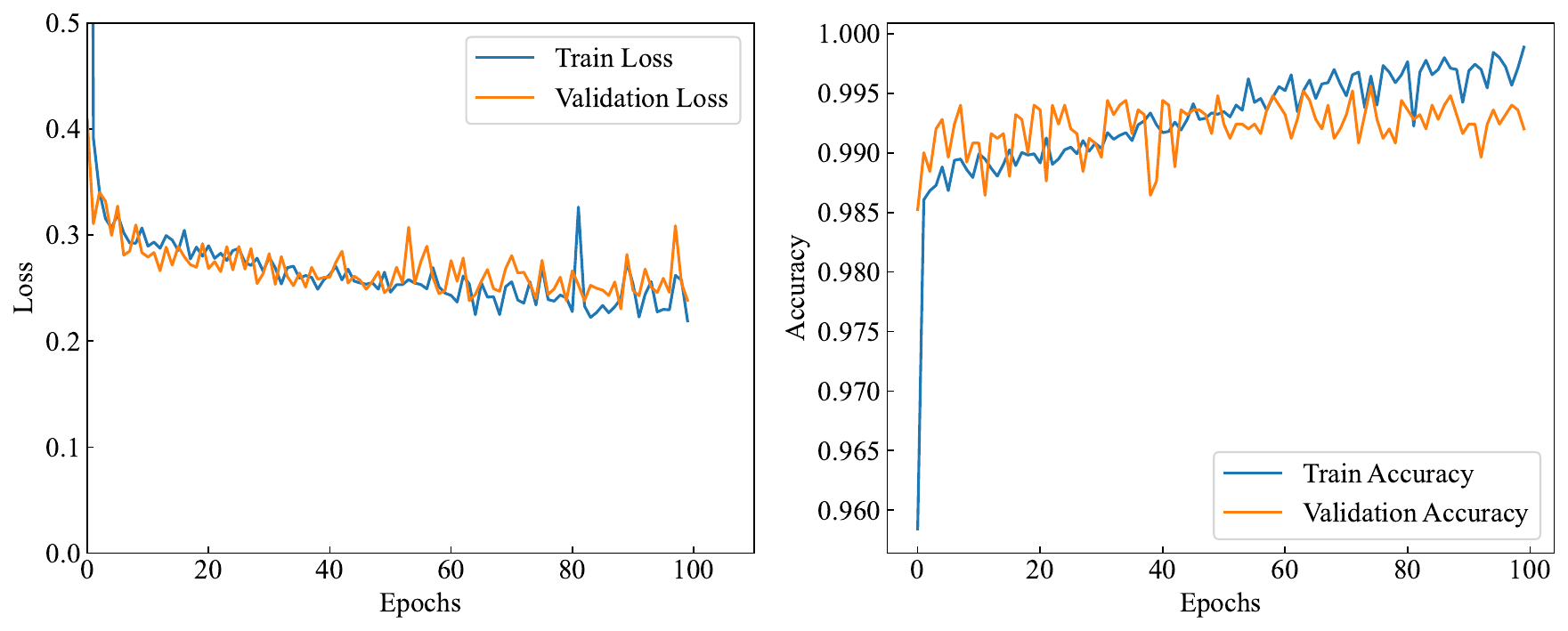}
\end{minipage}
\caption{Training and validation loss and accuracy of the model over 100 epochs. The blue and orange lines represent the training and validation sets, respectively. \label{fig4}}
\end{figure}
\begin{figure}[htbp]
\centering
\begin{minipage}[t]{0.5\textwidth}
\centering
\includegraphics[width=\textwidth]{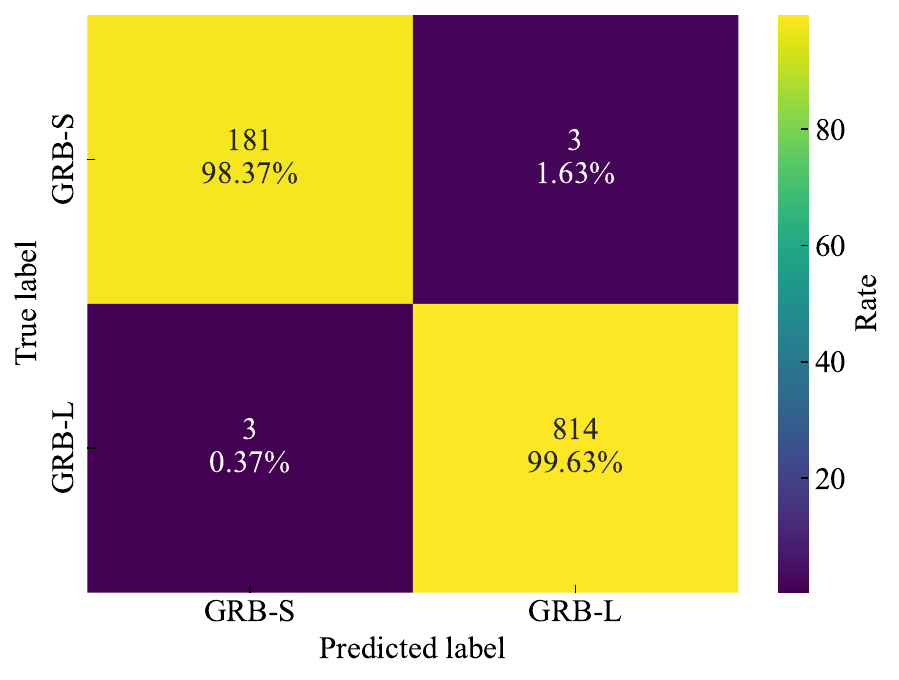}
\end{minipage}
\caption{Confusion matrix of the model's classification results on the test set. \label{fig5}}
\end{figure}
\begin{figure}[htbp]
\centering
\begin{minipage}[t]{0.4\textwidth}
\centering
\includegraphics[width=\textwidth]{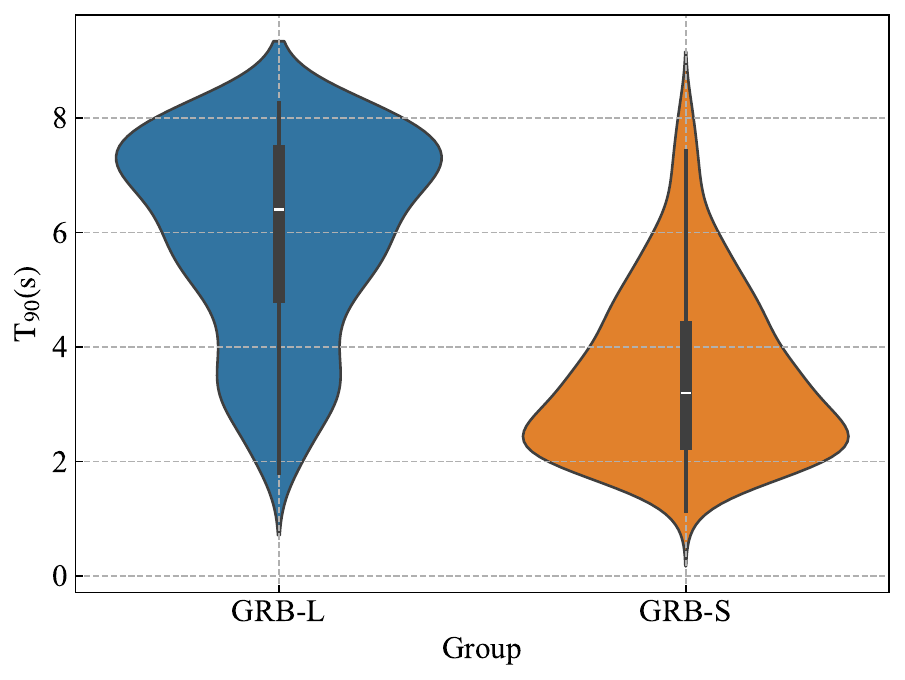}
\end{minipage}
\caption{Violin plot of the $T_\text{90}$ duration distribution for GRB-L and GRB-S. \label{fig6}}
\end{figure}

\begin{figure}[htbp]
\centering
\begin{minipage}[t]{0.4\textwidth}
\centering
\includegraphics[width=\textwidth]{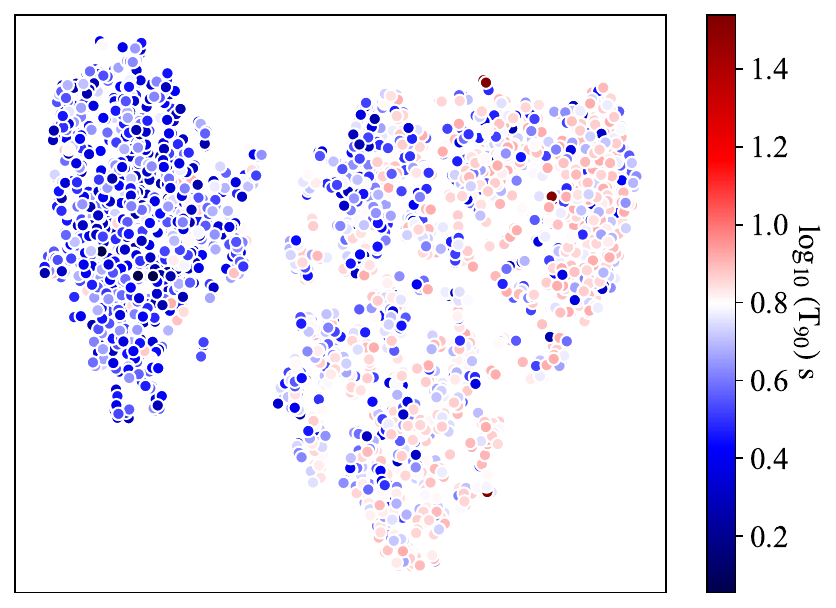}
\end{minipage}
\caption{Dimensionality reduction results of the extracted features from the count maps of overlapping GRBs. The color of the points represents the $log10(T_\text{90})$ values, ranging from blue (short duration) to red (long duration). \label{fig7}}
\end{figure}
\begin{figure}[htbp]
\centering
\subfigure[]
{
\begin{minipage}[t]{0.4\textwidth}
\centering
\includegraphics[width=\textwidth]{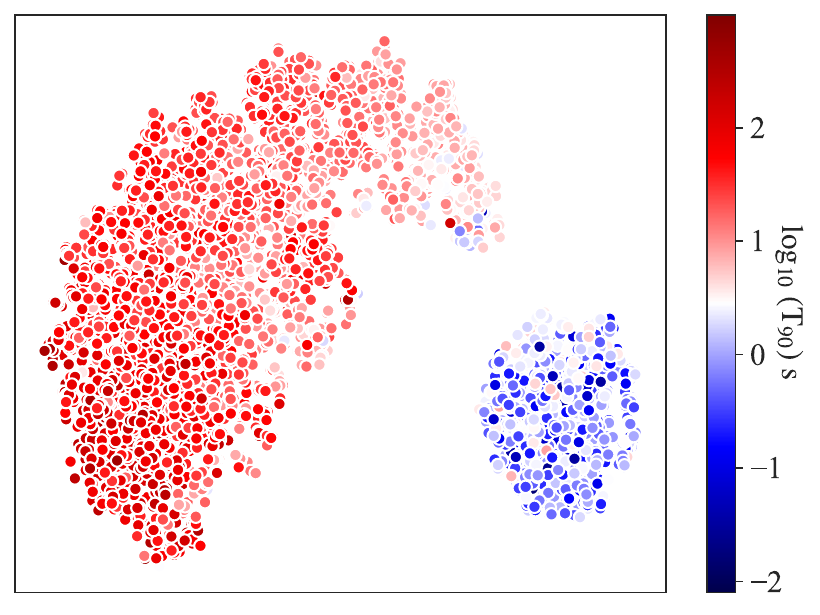}
\end{minipage}
}
\subfigure[]
{
\begin{minipage}[t]{0.5\textwidth}
\centering
\includegraphics[width=\textwidth]{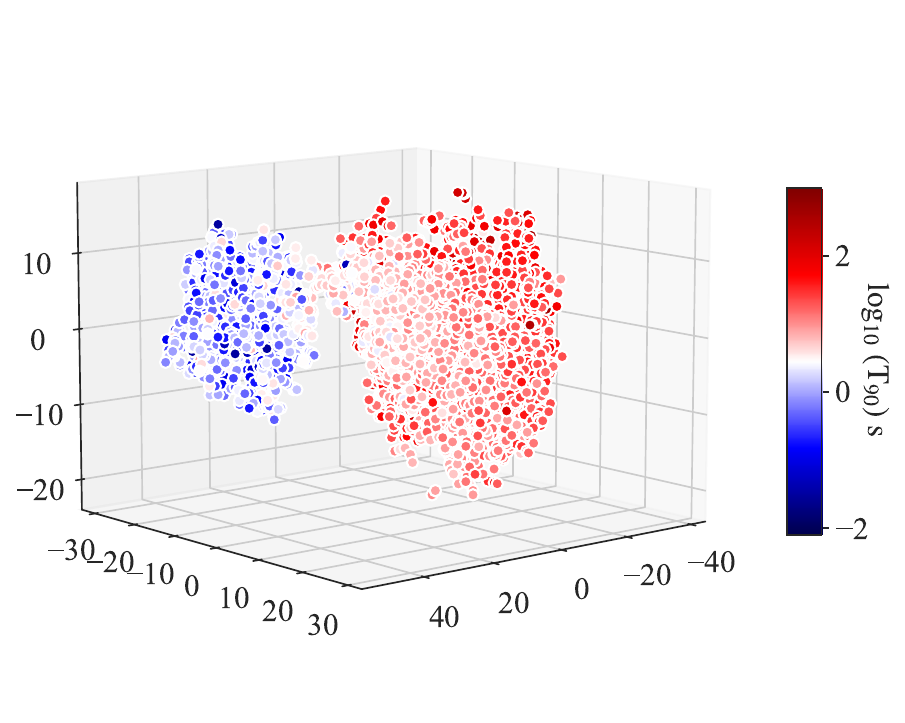}
\end{minipage}
}
\caption{2D and 3D t-SNE projections of GRB features extracted from count maps, colored by $\rm log_{10}(T_{90})$. \label{fig8}}
\end{figure}
\begin{figure}[htbp]
\centering
\begin{minipage}[t]{0.5\textwidth}
\centering
\includegraphics[width=\textwidth]{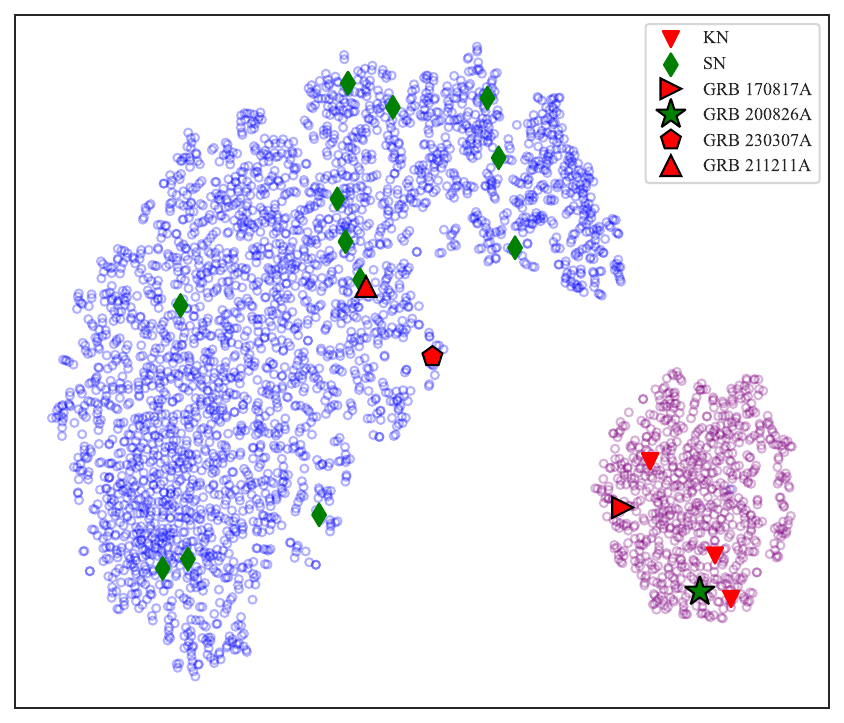}
\end{minipage}
\caption{Distribution of GRBs associated with KN and SN on the mapping of GRB count map feature extraction and dimensionality reduction. Pink and blue dots represent S- and L-type GRBs, respectively. GRBs associated with KN and SN are marked with red triangles and green diamonds, respectively.  \label{fig9}}
\end{figure}
\begin{figure}[htbp]
\centering
\begin{minipage}[t]{0.5\textwidth}
\centering
\includegraphics[width=\textwidth]{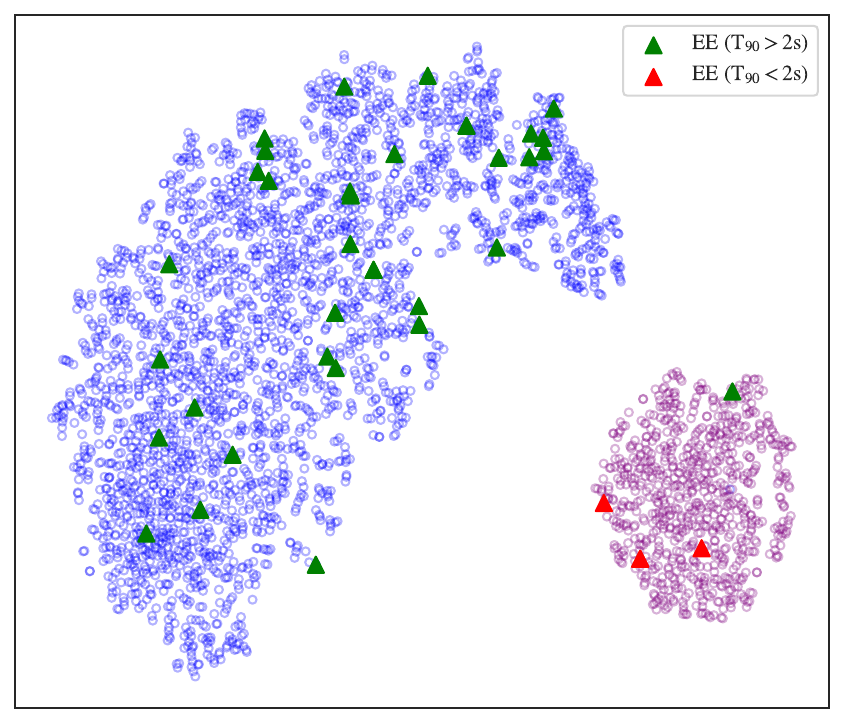}
\end{minipage}
\caption{Distribution of sGRB-EE in L- and S-type GRB samples. Red and green diamonds indicate $T_{90} < 2$ s and $T_{90} > 2$ s, respectively.  \label{fig10}}
\end{figure}

\begin{figure}[htbp]
\centering
\subfigure[]
{
\begin{minipage}[t]{0.45\textwidth}
\centering
\includegraphics[width=\textwidth]{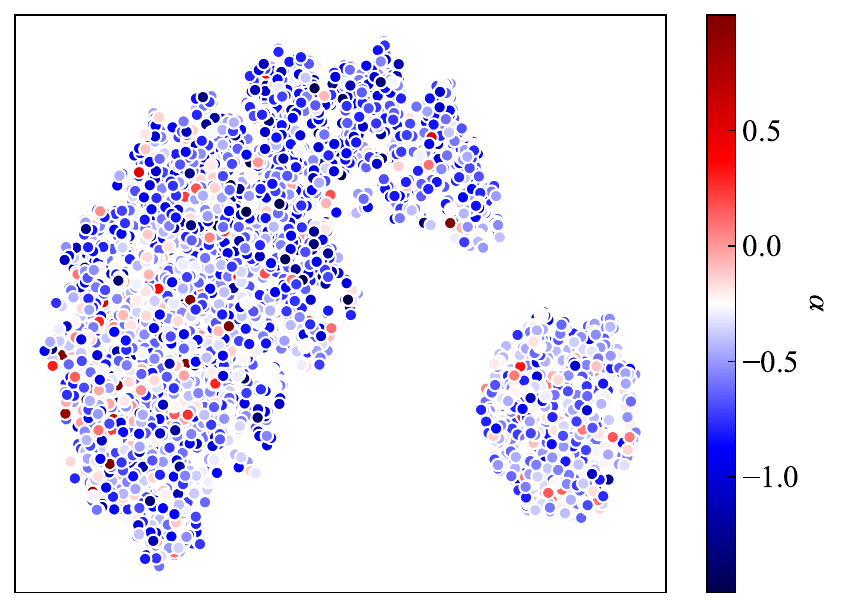}
\end{minipage}
}
\subfigure[]
{
\begin{minipage}[t]{0.45\textwidth}
\centering
\includegraphics[width=\textwidth]{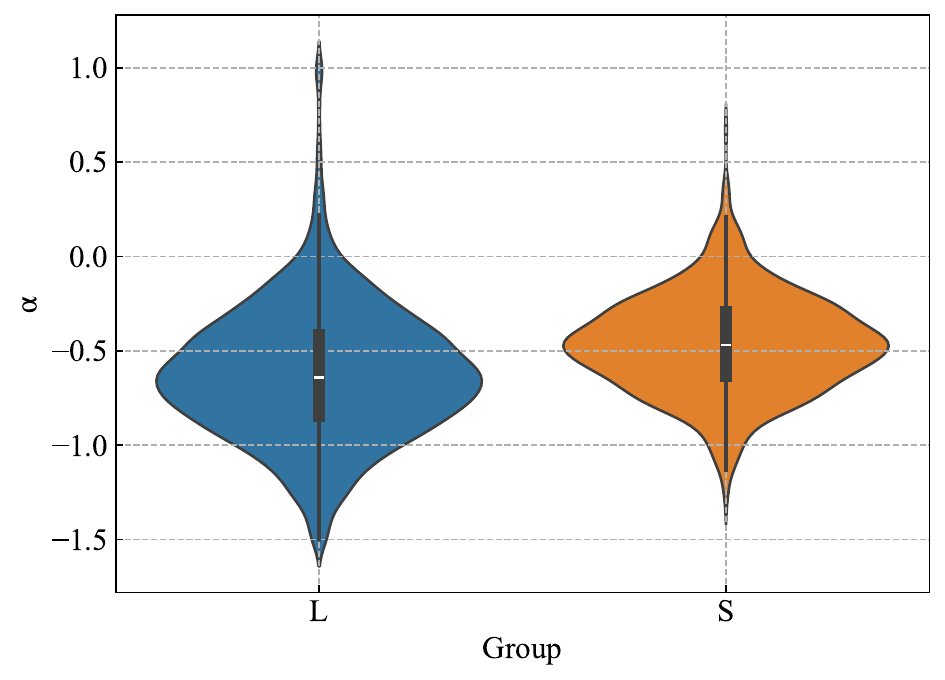}
\end{minipage}
}
\caption{Distribution of the $\alpha$ in L-type and S-type GRBs. Panel(a) each point represents an individual GRB event, with the color indicating the value of $\alpha$. The color bar ranges from -1.0 (blue) to 0.5 (red), showing the distribution of $\alpha$ values across events. Panel(b) The violin plot combines a box plot and a density plot to show the distribution shape and concentration trend of $\alpha$. The black box indicates the interquartile range, and the white dot represents the median. \label{fig11}}
\end{figure}
\begin{figure}[htbp]
\centering
\subfigure[]
{
\begin{minipage}[t]{0.45\textwidth}
\centering
\includegraphics[width=\textwidth]{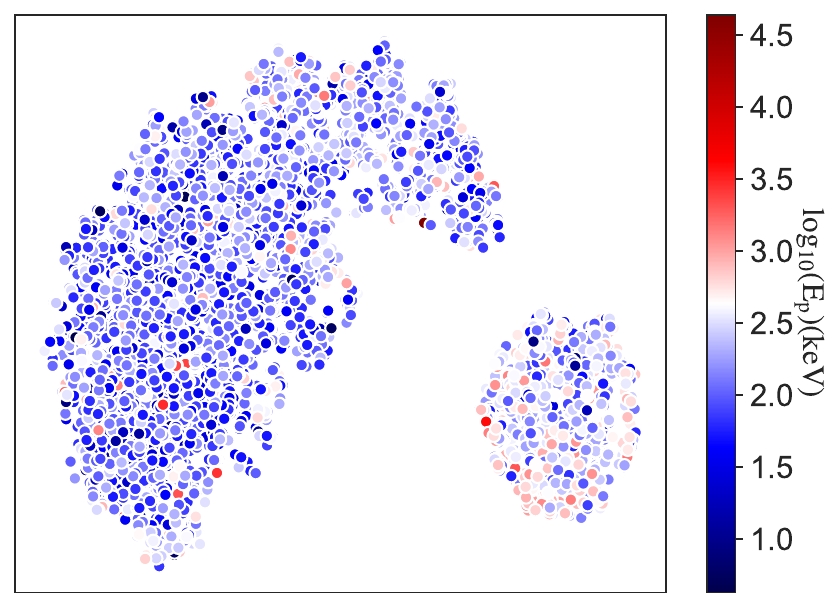}
\end{minipage}
}
\subfigure[]
{
\begin{minipage}[t]{0.45\textwidth}
\centering
\includegraphics[width=\textwidth]{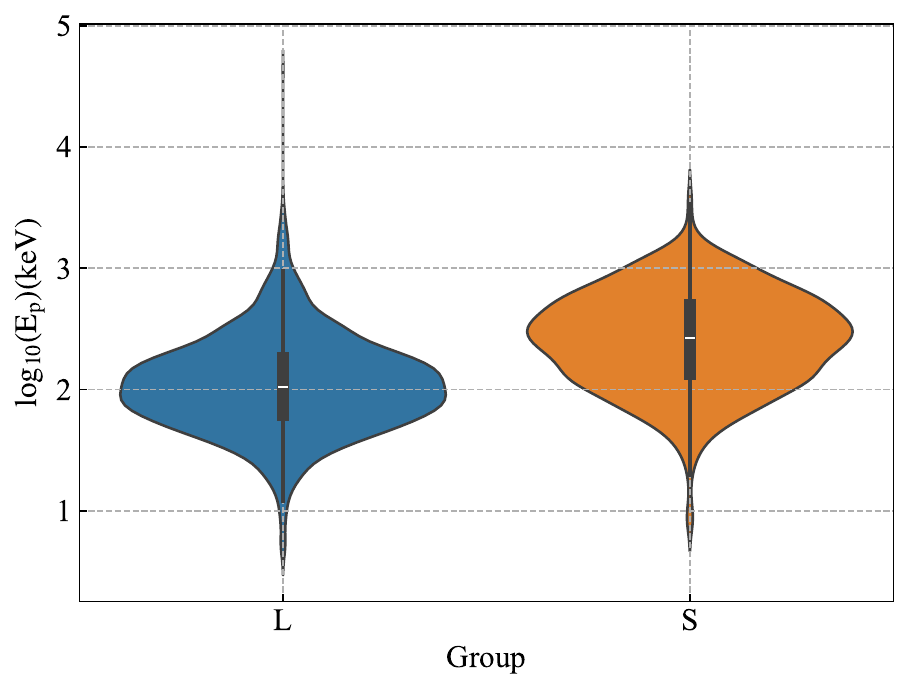}
\end{minipage}
}
\caption{Distribution of the $E_\text{p}$ in L-type and S-type GRBs. \label{fig12}}
\end{figure}
\begin{figure}[htbp]
\centering
\subfigure[]
{
\begin{minipage}[t]{0.45\textwidth}
\centering
\includegraphics[width=\textwidth]{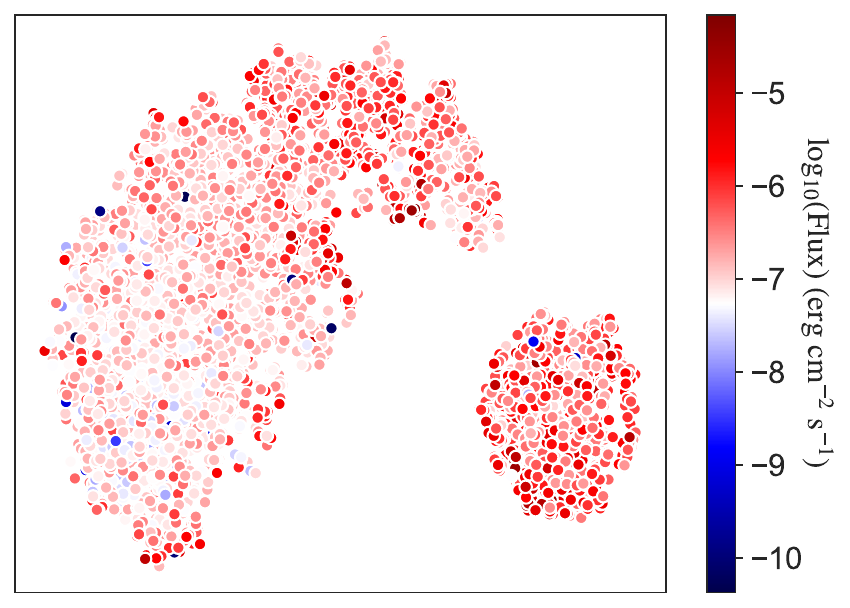}
\end{minipage}
}
\subfigure[]
{
\begin{minipage}[t]{0.45\textwidth}
\centering
\includegraphics[width=\textwidth]{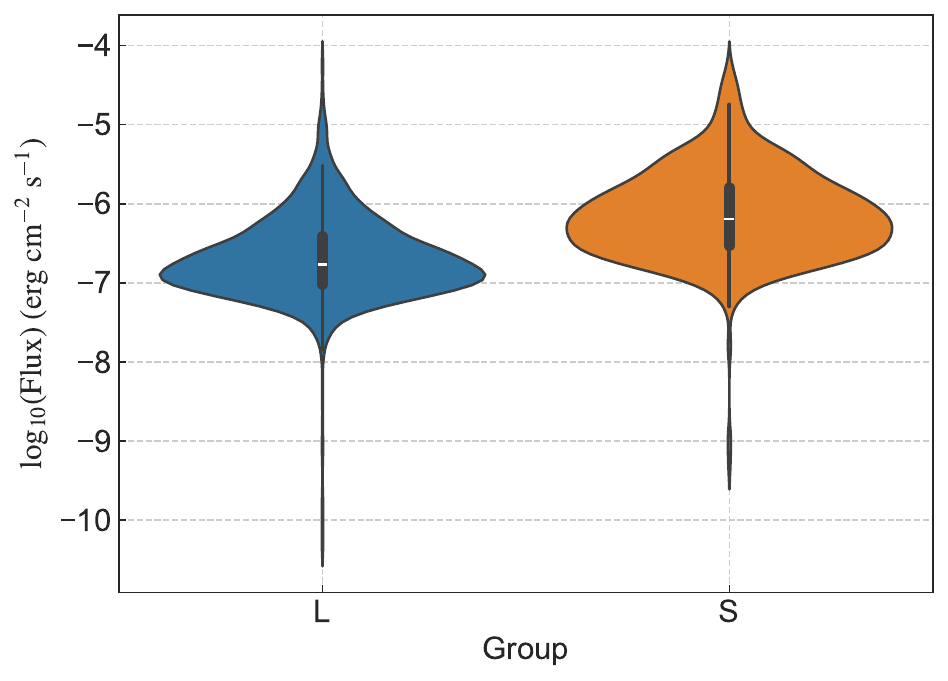}
\end{minipage}
}
\caption{Distribution of the Flux in L-type and S-type GRBs. \label{fig13}}
\end{figure}
\begin{figure}[htbp]
\centering
\subfigure[]
{
\begin{minipage}[t]{0.45\textwidth}
\centering
\includegraphics[width=\textwidth]{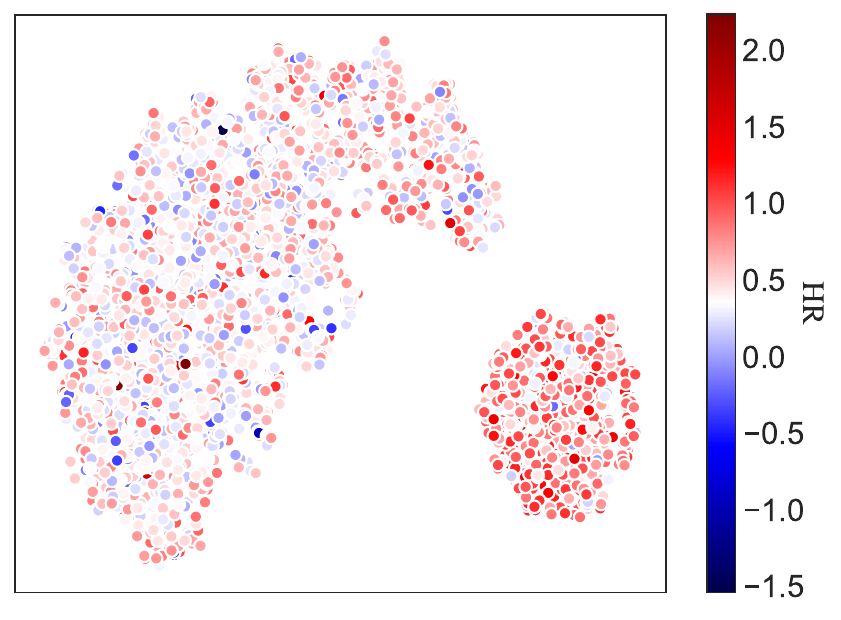}
\end{minipage}
}
\subfigure[]
{
\begin{minipage}[t]{0.45\textwidth}
\centering
\includegraphics[width=\textwidth]{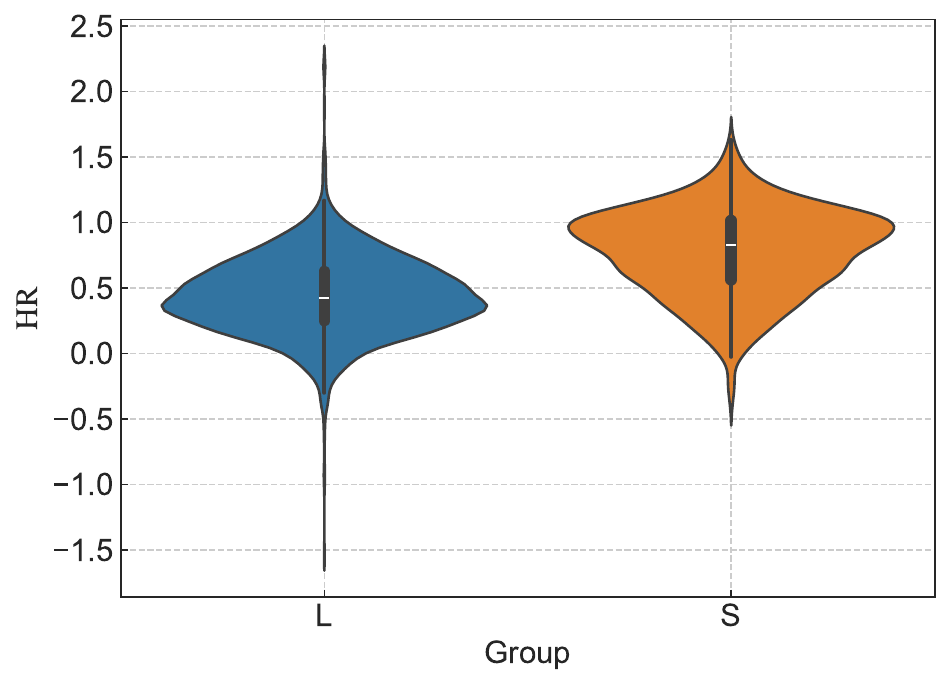}
\end{minipage}
}
\caption{Distribution of the HR in L-type and S-type GRBs. \label{fig14}}
\end{figure}
\begin{figure}[htbp]
\centering
\subfigure[]
{
\begin{minipage}[t]{0.45\textwidth}
\centering
\includegraphics[width=\textwidth]{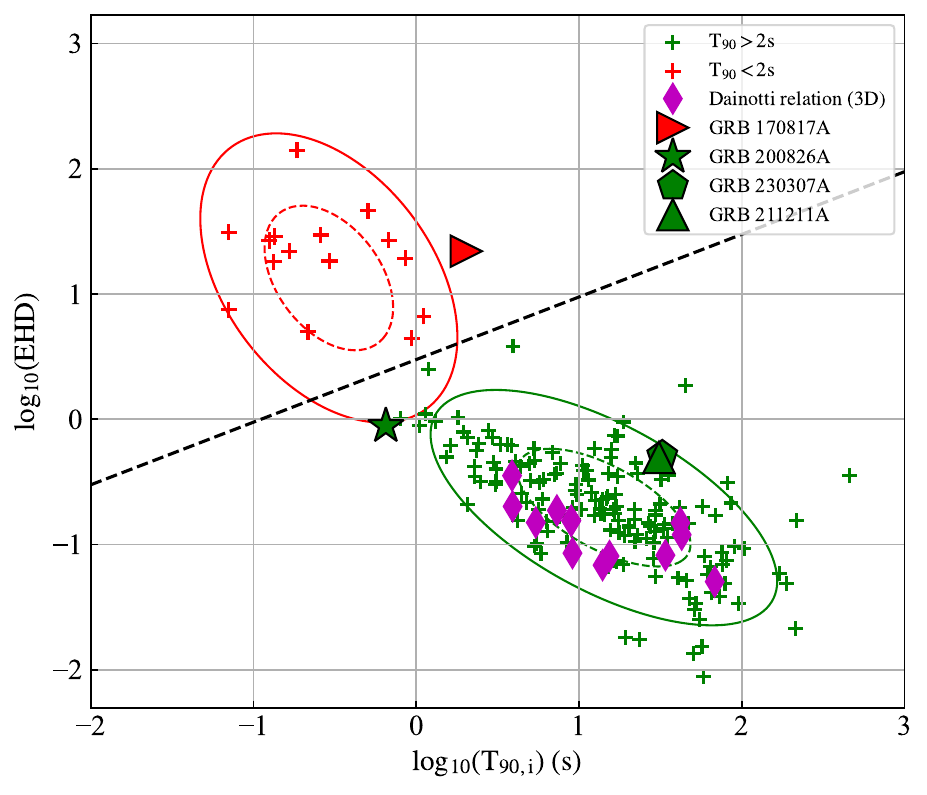}
\end{minipage}
}
\subfigure[]
{
\begin{minipage}[t]{0.45\textwidth}
\centering
\includegraphics[width=\textwidth]{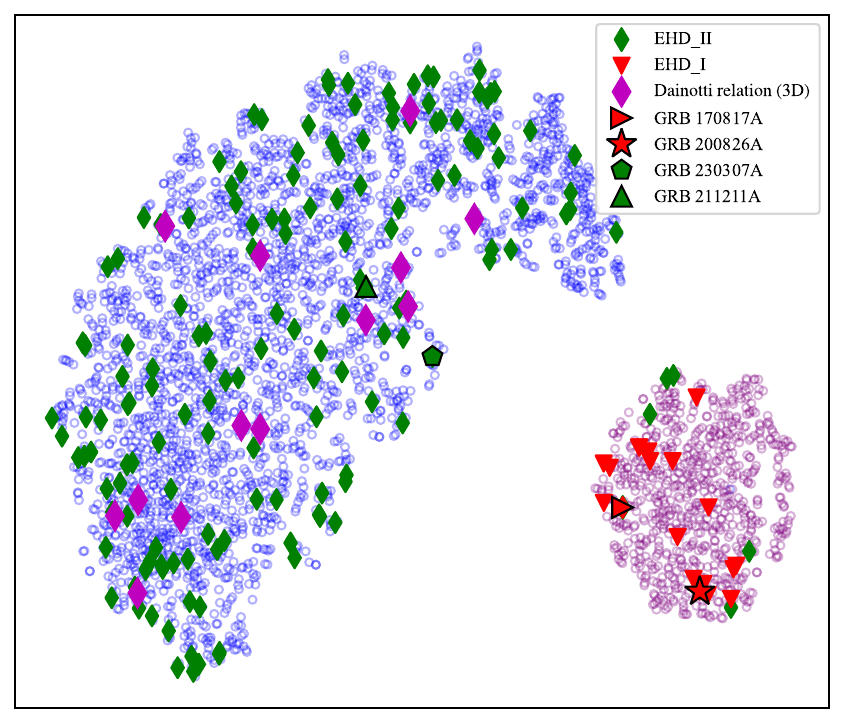}
\end{minipage}
}
\caption{Comparison of the results from the traditional classification method and our new classification method. Panel a: The relationship between the EHD parameter and $T_{90,i}$. The black dashed line serves as the classification boundary, clearly separating GRBs into Type I and Type II, with green and red symbols representing GRBs with $ T_{90} > 2$ s and $T_{90} < 2$ s, respectively. Panel b: Distribution of the classification results from the EHD-$T_{90,i}$ relation according to our new classification method. Green diamonds and red triangles represent GRBs classified as Type II ($\rm EHD\_II$) and Type I ($\rm EHD\_I$) under the EHD parameter classification method, respectively. Special GRBs, including GRB 200826A, GRB 230307A, GRB 211211A, and GRB 170817A, are highlighted with unique symbols. The magenta diamond marker represents a platinum sample that closely follows the Dainotti relation (3D). \label{fig15}}

\end{figure}
\begin{figure}[htbp]
\centering
\begin{minipage}[t]{0.5\textwidth}
\centering
\includegraphics[width=\textwidth]{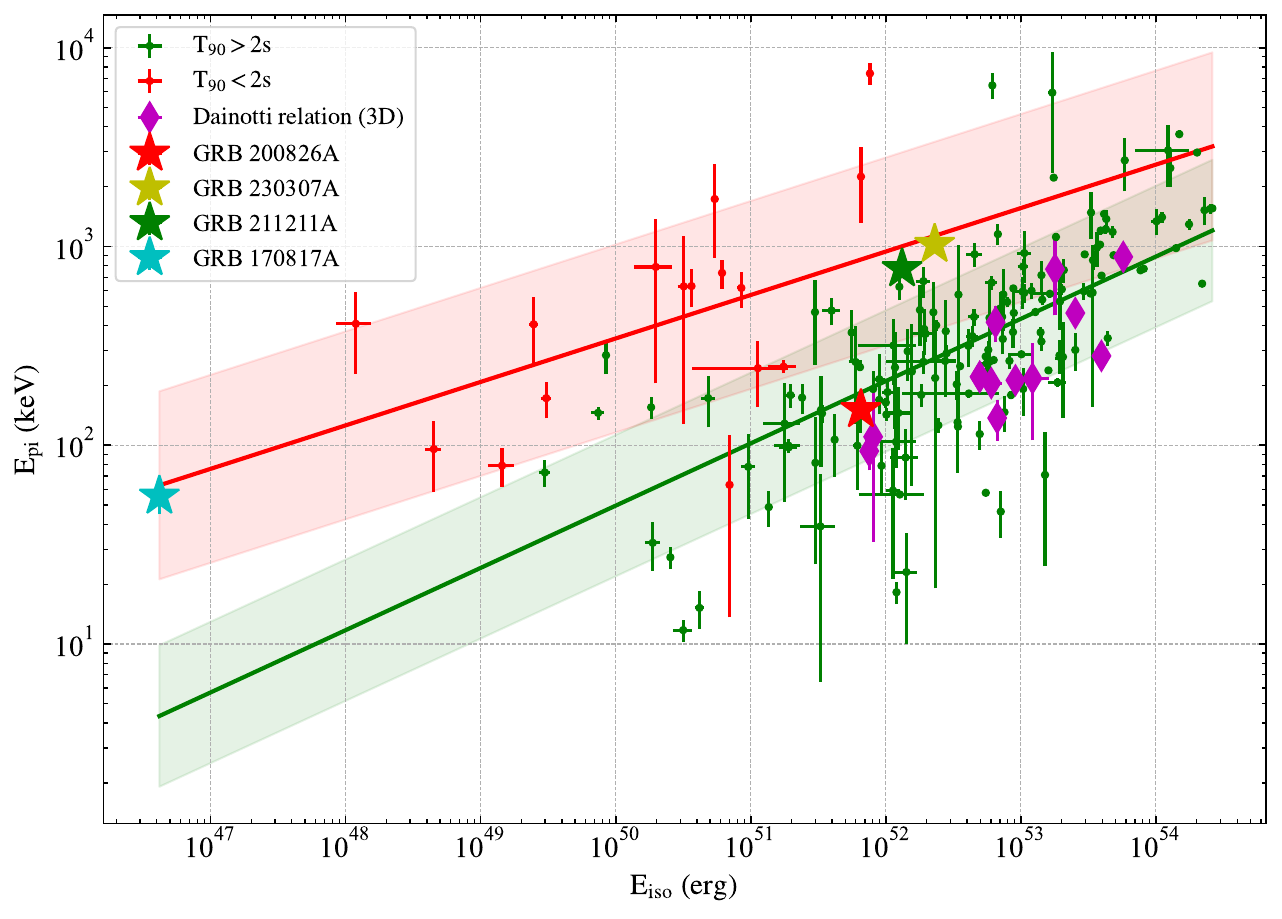}
\end{minipage}
\caption{The relationship between $E_\text{p,i}$ and isotropic equivalent energy $E_{\text{iso}}$ (Amati relation). Green symbol denote GRBs with $T_\text{90} > 2$ s, while red symbol denote GRBs with $T_\text{90} < 2$ s. Specific GRBs, including GRB 200826A, GRB 230307A, GRB 211211A, and GRB 170817A, are highlighted with star markers. The green line indicates the Amati relation for LGRBs, and the red line suggests a possible relation for SGRBs. The magenta diamond marker represents  Platinum sample that closely adheres to the Dainotti relation (3D). The shaded areas around the lines represent the 1$\sigma$ confidence intervals. \label{fig16}}
\end{figure}

\begin{figure}[htbp]
\centering
\begin{minipage}[t]{0.5\textwidth}
\centering
\includegraphics[width=\textwidth]{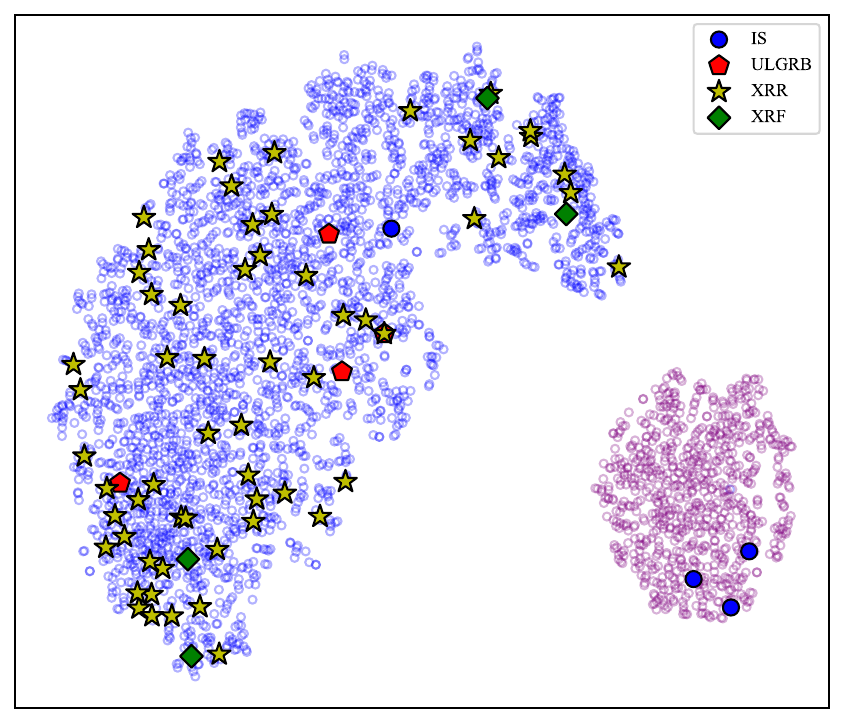}
\end{minipage}
\caption{Comparison with the unsupervised machine learning classification results of \cite{2023MNRAS.525.5204B}. The blue, red, yellow, and green symbols represent intrinsically short GRBs (IS), ultra-long GRBs (ULGRB), X-ray rich (XRR), and X-ray flash (XRF), respectively.  \label{fig17}}
\end{figure}

\begin{table}[ht]
\centering
\caption{Performance Metrics for Different Data Sets}
\begin{tabular}{lcccc}
\hline
\hline
Data Set    & Accuracy (\%) & Precision (\%) &Recall (\%) & F1-score (\%) \\
\hline
Training set         & 99.72                  & 99.79                   & 99.86                & 99.83                  \\
Validation set       & 99.56                  & 99.76                   & 99.71                & 99.73                  \\
Test set             & 99.40                  & 99.63                   & 99.63                & 99.63                  \\
\hline
\end{tabular}
\label{tab1}
\end{table}

\begin{table}[!ht]
    \centering
    \caption{Classification results of GRBs with overlapping $T_\text{90}$.}
    \begin{threeparttable} 
    \begin{tabular}{ccc}
    \hline
    \hline
    GRB & $T_\text{90}$ (s) & Class \\
    \hline
       GRB080714086 & 5.376 & GRB-L \\ 
        GRB080715950 & 7.872 &GRB-L \\ 
        GRB080806584 & 2.304 &GRB-S \\
        GRB080808451 & 4.352 &GRB-S \\
        GRB080816989 & 4.608 &GRB-S \\
        GRB080817720 & 4.416 &GRB-S \\
        GRB080821332 & 5.888 &GRB-S \\
        GRB080824909 & 7.424 &GRB-L \\
        GRB080828189 & 3.008 &GRB-L \\
        GRB080829790 & 7.68 &GRB-S \\
        GRB080906212 & 2.875 &GRB-L \\
        GRB081006604 & 6.4 &GRB-S \\
        GRB081006872 & 3.328 &GRB-S \\
        GRB081101532 & 8.256 &GRB-L \\
        GRB081130212 & 2.24 &GRB-S \\
        GRB081204004 & 7.424 &GRB-L \\
        GRB081206604 & 7.936 &GRB-L \\
        GRB081215784 & 5.568 &GRB-L \\
        GRB081215880 & 7.68 &GRB-L \\
        GRB090120627 & 1.856 &GRB-S \\
        GRB090126227 & 5.632 &GRB-S \\
        GRB090225009 & 2.176 &GRB-L \\
        GRB090228976 & 7.936 &GRB-L \\
        GRB090304216 & 2.816 &GRB-S \\
        GRB090305052 & 1.856 &GRB-S \\
        GRB090320045 & 2.368 &GRB-S \\
        GRB090320418 & 7.936 &GRB-L \\
    \hline
    \end{tabular}
    \tablecomments{A full machine-readable version is available online.}
    \end{threeparttable}  
    \label{tab2}
\end{table}

\begin{table}[!ht]
    \centering
    \caption{sGRB-EE Sample}
    \begin{tabular}{ll|l l}
    \hline
    \hline
        GRB & $T_\text{90}$ (s) & GRB & $T_\text{90}$ (s)\\ \hline
        GRB080807993 & 19.072 & GRB120605453 & 18.112 \\
        GRB081110601 & 11.776 & GRB130628531 & 21.504 \\ 
        GRB081110601 & 11.776 & GRB131108862 & 18.176 \\ 
        GRB081129161 & 62.657 & GRB140308710 & 12.032 \\ 
        GRB081215784 & 5.568 & GRB140819160 & 6.656 \\
        GRB090227772 & 0.304 & GRB141229492 & 13.824 \\ 
        GRB090510016 & 0.96 & GRB150127398 & 52.736 \\
        GRB090720710 & 10.752 & GRB150510139 & 51.904 \\
        GRB090831317 & 39.424 & GRB150702998 & 45.825 \\ 
        GRB090831317 & 39.424 & GRB160721806 & 9.984 \\
        GRB090929190 & 6.174 & GRB161218356 & 25.857 \\
        GRB091127976 & 8.701 & GRB170527480 & 49.153 \\
        GRB100829876 & 8.704 & GRB170626401 & 12.288 \\
        GRB100916779 & 12.8 & GRB170728961 & 46.336 \\
        GRB110824009 & 76.607 & GRB180618030 & 3.712 \\
        GRB111012811 & 7.936 & GRB190308923 & 45.568 \\
        GRB111221739 & 27.136 & GRB200219317 & 1.152 \\
        GRB120119229 & 41.728 & GRB200313456 & 5.184 \\
        GRB120304248 & 5.376 & GRB201104001 & 52.48 \\
    \hline
    \end{tabular}
\label{tab3}
\end{table}

\begin{table}[!ht]
    \centering
    \caption{Time-integrated spectrum analysis results of 3726 GRBs and their classification results.}
    \begin{tabular}{cccccccc}
    \hline
    \hline
   GRB & $t_\text{start}-t_\text{end} $ & $\alpha$ & $\beta$ &$E_\text{p}$&Flux &HR&Class\\
 & s& & &$keV$& $\times10^{-6}erg^{-1}cm^{-2}s^{-1}$& & \\
    \hline
GRB080714086&-3.97$-$0.32&${-0.35}^{+0.19}_{-0.38}$ &${-2.14}^{+0.46}_{-0.25}$ &${256.8}^{+78.45}_{-53.47}$& ${0.2}^{+0.02}_{-0.03}$&7.64&L\\
GRB080714425&-5.63$-$5.38&${-0.61}^{+0.06}_{-0.45}$ &${-1.71}^{+0.22}_{-0.05}$ &${22.74}^{+3.77}_{-15.17}$& ${0.06}^{+0.01}_{-0.01}$&1.84&L\\
GRB080714745&-1.54$-$9.98&${-0.49}^{+0.24}_{-0.37}$ &${-2.01}^{+0.26}_{-0.08}$ &${57.62}^{+9.85}_{-15.95}$& ${0.1}^{+0.01}_{-0.01}$&1.81&L\\
GRB080715950&-6.91$-$6.14&${-1.13}^{+0.07}_{-0.07}$ &${-2.43}^{+0.39}_{-0.23}$ &${269.93}^{+41.57}_{-50.14}$& ${0.61}^{+0.03}_{-0.03}$&3.23&L\\
GRB080717543&-2.05$-$9.98&${-0.97}^{+0.09}_{-0.65}$ &${-2.34}^{+0.19}_{-0.47}$ &${103.98}^{+51.93}_{-9.54}$& ${0.11}^{+0.01}_{-0.01}$&2.08&L\\
GRB080719529&-1.02$-$8.19&${-0.31}^{+0.28}_{-0.37}$ &${-1.97}^{+0.48}_{-0.13}$ &${61.87}^{+12.99}_{-20.4}$& ${0.07}^{+0.01}_{-0.01}$&2.07&L\\
GRB080723557&-5.12$-$3.84&${-0.51}^{+0.13}_{-0.23}$ &${-1.94}^{+0.05}_{-0.05}$ &${273.59}^{+46.01}_{-40.37}$& ${1.11}^{+0.04}_{-0.04}$&6.68&L\\
GRB080723913&-2.05$-$2.3&${-0.43}^{+0.37}_{-0.34}$ &${-1.74}^{+0.55}_{-0.02}$ &${153.14}^{+25.76}_{-149.3}$& ${0.84}^{+0.13}_{-0.13}$&4.68&S\\
GRB080723985&-0.06$-$0.26&${-0.94}^{+0.03}_{-0.04}$ &${-2.73}^{+0.42}_{-0.13}$ &${431.54}^{+26.59}_{-39.02}$& ${0.77}^{+0.02}_{-0.02}$&4.89&L\\
GRB080724401&1.86$-$36.42&${-0.88}^{+0.11}_{-0.11}$ &${-2.48}^{+0.36}_{-0.17}$ &${107.09}^{+10.45}_{-11.84}$& ${0.28}^{+0.02}_{-0.02}$&2.22&L\\
GRB080725435&0.32$-$21.31&${-1.03}^{+0.07}_{-0.06}$ &${-2.59}^{+0.36}_{-0.26}$ &${309.92}^{+35.61}_{-58.36}$& ${0.32}^{+0.01}_{-0.02}$&3.87&L\\
GRB080725541&-6.34$-$3.65&${-0.6}^{+0.17}_{-0.22}$ &${-2.21}^{+0.46}_{-0.2}$ &${790.6}^{+210.72}_{-383.72}$& ${0.9}^{+0.08}_{-0.08}$&9.64&S\\
GRB080727964&-10.5$-$1.79&${-0.94}^{+0.14}_{-0.17}$ &${-2.23}^{+0.47}_{-0.16}$ &${168.59}^{+26.3}_{-39.59}$& ${0.15}^{+0.01}_{-0.01}$&3.08&L\\
GRB080730520&-2.05$-$7.17&${-0.77}^{+0.13}_{-0.12}$ &${-2.34}^{+0.33}_{-0.1}$ &${131.23}^{+9.51}_{-22.71}$& ${0.3}^{+0.01}_{-0.02}$&2.89&L\\
GRB080730786&-17.92$-$63.23&${-0.68}^{+0.07}_{-0.07}$ &${-2.94}^{+0.3}_{-0.16}$ &${124.08}^{+5.42}_{-6.81}$& ${0.45}^{+0.02}_{-0.02}$&2.82&L\\
GRB080802386&-0.64$-$0.19&${-0.54}^{+0.23}_{-0.26}$ &${-2.06}^{+0.55}_{-0.16}$ &${369.03}^{+84.15}_{-215.8}$& ${1.1}^{+0.12}_{-0.13}$&7.85&S\\
GRB080803772&-13.57$-$34.3&${-0.29}^{+0.13}_{-0.26}$ &${-2.53}^{+0.34}_{-0.34}$ &${253.02}^{+47.47}_{-26.72}$& ${0.19}^{+0.01}_{-0.01}$&8.02&L\\
GRB080804456&-1.6$-$1.28&${-0.47}^{+0.2}_{-0.3}$ &${-2.48}^{+0.44}_{-0.25}$ &${131.86}^{+20.34}_{-19.39}$& ${0.1}^{+0.01}_{-0.01}$&3.53&L\\
GRB080804972&-5.89$-$24.58&${-0.64}^{+0.07}_{-0.11}$ &${-2.42}^{+0.37}_{-0.17}$ &${242.28}^{+24.28}_{-26.38}$& ${0.38}^{+0.02}_{-0.02}$&5.33&L\\
GRB080805496&0.58$-$27.2&${-0.82}^{+0.11}_{-0.39}$ &${-2.43}^{+0.25}_{-0.11}$ &${22.68}^{+2.48}_{-4.76}$& ${0.06}^{+0.01}_{-0.01}$&0.66&L\\
GRB080805584&-9.73$-$9.73&${-0.72}^{+0.22}_{-0.42}$ &${-1.98}^{+0.43}_{-0.13}$ &${49.99}^{+16.79}_{-24.85}$& ${0.06}^{+0.01}_{-0.01}$&1.58&L\\
GRB080806584&8.7$-$34.56&${-0.29}^{+0.25}_{-0.28}$ &${-2.46}^{+0.35}_{-0.21}$ &${69.77}^{+9.67}_{-10.11}$& ${0.17}^{+0.02}_{-0.02}$&1.84&S\\
GRB080807993&0.48$-$6.3&${-0.92}^{+0.1}_{-0.1}$ &${-1.93}^{+0.6}_{-0.07}$ &${500.54}^{+112.81}_{-270.0}$& ${0.36}^{+0.03}_{-0.03}$&5.29&L\\
GRB080808451&-0.26$-$0.26&${-0.25}^{+0.18}_{-0.4}$ &${-2.28}^{+0.4}_{-0.26}$ &${117.89}^{+23.34}_{-15.4}$& ${0.16}^{+0.02}_{-0.02}$&3.7&S\\
GRB080808565&-0.13$-$0.45&${-0.76}^{+0.08}_{-0.17}$ &${-2.88}^{+0.32}_{-0.2}$ &${65.4}^{+5.13}_{-3.36}$& ${0.21}^{+0.01}_{-0.01}$&1.33&L\\
GRB080808772&-7.94$-$36.35&${-0.45}^{+0.02}_{-0.53}$ &${-2.13}^{+0.52}_{-0.29}$ &${18.24}^{+3.02}_{-20.95}$& ${0.02}^{+0.01}_{-0.01}$&0.94&L\\
GRB080809808&-13.06$-$16.13&${-0.9}^{+0.15}_{-0.5}$ &${-2.45}^{+0.29}_{-0.33}$ &${58.08}^{+17.12}_{-7.82}$& ${0.15}^{+0.02}_{-0.02}$&1.27&L\\
GRB080810549&-8.45$-$3.58&${-0.6}^{+0.19}_{-0.2}$ &${-2.22}^{+0.46}_{-0.18}$ &${327.0}^{+56.09}_{-110.52}$& ${0.21}^{+0.01}_{-0.02}$&6.72&L\\
GRB080812889&-4.35$-$35.84&${-0.11}^{+0.19}_{-0.27}$ &${-2.3}^{+0.45}_{-0.16}$ &${146.7}^{+19.89}_{-19.99}$& ${0.18}^{+0.02}_{-0.02}$&5.24&L\\
GRB080815917&-2.82$-$2.56&${-0.63}^{+0.26}_{-0.3}$ &${-1.79}^{+0.45}_{-0.1}$ &${169.16}^{+39.28}_{-88.51}$& ${0.63}^{+0.08}_{-0.08}$&4.26&S\\
GRB080816503&-9.73$-$52.74&${-0.9}^{+0.07}_{-0.1}$ &${-2.53}^{+0.34}_{-0.16}$ &${125.68}^{+11.33}_{-9.75}$& ${0.21}^{+0.01}_{-0.01}$&2.5&L\\
GRB080816989&0.45$-$16.51&${-0.55}^{+0.13}_{-0.1}$ &${-2.43}^{+0.31}_{-0.28}$ &${1443.9}^{+329.11}_{-359.53}$& ${0.69}^{+0.04}_{-0.04}$&11.84&S\\
GRB080817161&5.06$-$116.48&${-0.96}^{+0.03}_{-0.03}$ &${-2.17}^{+0.22}_{-0.1}$ &${397.17}^{+31.99}_{-43.21}$& ${0.84}^{+0.02}_{-0.02}$&4.68&L\\
GRB080817720&-0.51$-$22.53&${-0.61}^{+0.16}_{-0.21}$ &${-2.26}^{+0.46}_{-0.31}$ &${687.47}^{+234.92}_{-240.45}$& ${0.45}^{+0.05}_{-0.04}$&9.07&S\\
GRB080818579&-0.26$-$23.3&${-0.5}^{+0.16}_{-0.44}$ &${-1.93}^{+0.34}_{-0.15}$ &${69.08}^{+22.04}_{-17.82}$& ${0.07}^{+0.01}_{-0.01}$&2.18&L\\
GRB080818945&-0.83$-$2.43&${-0.97}^{+0.2}_{-0.28}$ &${-2.49}^{+0.34}_{-0.16}$ &${56.58}^{+7.94}_{-8.51}$& ${0.14}^{+0.01}_{-0.01}$&1.2&L\\
GRB080821332&-0.26$-$80.13&${-0.84}^{+0.11}_{-0.09}$ &${-2.54}^{+0.36}_{-0.15}$ &${109.22}^{+7.35}_{-13.41}$& ${0.6}^{+0.03}_{-0.03}$&2.3&L\\
GRB080823363&-1.54$-$6.66&${-1.2}^{+0.12}_{-0.23}$ &${-2.12}^{+0.41}_{-0.19}$ &${147.94}^{+46.28}_{-52.19}$& ${0.15}^{+0.01}_{-0.01}$&2.27&L\\
GRB080824909&-1.02$-$43.01&${-0.89}^{+0.1}_{-0.15}$ &${-2.4}^{+0.41}_{-0.19}$ &${132.39}^{+17.48}_{-21.35}$& ${0.42}^{+0.02}_{-0.03}$&2.64&L\\
GRB080825593&2.56$-$143.36&${-0.6}^{+0.05}_{-0.03}$ &${-2.39}^{+0.2}_{-0.08}$ &${180.41}^{+5.23}_{-10.9}$& ${1.6}^{+0.03}_{-0.03}$&4.35&L\\
GRB080828189&-0.06$-$0.35&${-0.39}^{+0.3}_{-0.35}$ &${-2.23}^{+0.51}_{-0.21}$ &${182.12}^{+45.54}_{-73.71}$& ${0.15}^{+0.02}_{-0.03}$&5.35&L\\
GRB080829790&-1.54$-$13.82&${-0.38}^{+0.14}_{-0.24}$ &${-2.64}^{+0.23}_{-0.16}$ &${69.18}^{+5.68}_{-4.78}$& ${0.28}^{+0.02}_{-0.02}$&1.67&S\\
GRB080830368&-0.64$-$0.06&${-0.84}^{+0.1}_{-0.24}$ &${-2.24}^{+0.43}_{-0.26}$ &${171.33}^{+40.54}_{-31.76}$& ${0.15}^{+0.01}_{-0.01}$&3.38&L\\
GRB080831053&-1.79$-$20.48&${-0.51}^{+0.14}_{-0.53}$ &${-2.11}^{+0.51}_{-0.21}$ &${118.26}^{+45.4}_{-37.94}$& ${0.14}^{+0.04}_{-0.04}$&3.23&S\\
GRB080831921&0.32$-$9.79&${-0.57}^{+0.16}_{-0.34}$ &${-2.43}^{+0.33}_{-0.19}$ &${77.94}^{+12.51}_{-7.56}$& ${0.11}^{+0.01}_{-0.01}$&1.9&L\\
GRB080904886&-3.84$-$16.38&${-0.93}^{+0.13}_{-0.23}$ &${-2.72}^{+0.13}_{-0.14}$ &${34.76}^{+2.75}_{-1.85}$& ${0.29}^{+0.01}_{-0.01}$&0.69&L\\
GRB080905499&-10.24$-$8.96&${0.05}^{+0.32}_{-0.25}$ &${-2.1}^{+0.52}_{-0.16}$ &${357.45}^{+49.96}_{-145.47}$& ${0.76}^{+0.08}_{-0.08}$&15.95&S\\
    \hline
    \end{tabular}
    \label{tab4}
    \tablecomments{A full machine-readable version is available online.}
\end{table}

\begin{table}[!ht]
    \centering
    \caption{Results of the 2D KS test comparing the spectral characteristics of L-type and S-type GRBs.}
    \begin{tabular}{ccc}
    \hline
     Parameter & KS& p-Value \\
    \hline
    $\alpha$ & 0.26 & $p<0.0001$ \\
    $E_\text{p}$ & 0.40 & $p<0.0001$ \\ 
    Flux & 0.45 & $p<0.0001$ \\ 
    HR & 0.45 & $p<0.0001$ \\
    \hline
    \end{tabular}
    \label{tab5}
\end{table}

\end{document}